\documentclass[11pt]{article}

\usepackage{amsmath,amsfonts,amssymb,amsthm,bbm,stmaryrd}
\usepackage[dvips]{graphicx}
\usepackage{latexsym}
\usepackage{psfrag}
\usepackage[latin1]{inputenc}
\usepackage{hyperref}
\usepackage{color}
\usepackage{enumerate}
\numberwithin{equation}{section}
\usepackage{float}
\usepackage{cite}

\newcommand{\be}{\begin{equation}}
\newcommand{\ee}{\end{equation}}
\newcommand{\barray}{\begin{array}}
\newcommand{\earray}{\end{array}}
\newcommand{\bea}{\begin{eqnarray}}
\newcommand{\eea}{\end{eqnarray}}
\newcommand{\bs}{\begin{subequations}}
\newcommand{\es}{\end{subequations}}

\newcommand{\bit}{\begin{itemize}}
\newcommand{\eit}{\end{itemize}}
\newcommand{\bd}{\begin{description}}
\newcommand{\ed}{\end{description}}

\def\nn{\nonumber}

\newcommand{\p}{\partial}

\newcommand{\f}{\frac}

  \newcommand{\g}{\gamma}  
\renewcommand{\d}{\delta}  \newcommand{\eps}{\epsilon}

\let\m=\mu    \let\n=\nu    
       
 \let\D=\Delta

\def\cN{{\cal N}}

\newcommand{\pbi}[1]{\underset{\leftarrow}{#1}}

\begin{document}

\title{\bf Second law from the Noether current on null hypersurfaces}

\author{\Large{Antoine Rignon-Bret}
\smallskip \\ 
\small{\it{Aix Marseille Univ., Univ. de Toulon, CNRS, CPT, UMR 7332, 13288 Marseille, France}} }

\maketitle

\begin{abstract}
    I study the balance law equation of surface charges in the presence of background fields. The construction allows a unified description of Noether's theorem for both global and local symmetries. From the balance law associated with some of these symmetries, I will discuss generalizations of Wald's Noether entropy formula and general entropy balance laws on null hypersurfaces based on the null energy conditions, interpreted as an entropy creation term. The entropy is generally the so-called improved Noether charge, a quantity that has recently been investigated by many authors, associated to null future-pointing diffeomorphisms. These local and dynamical definitions of entropy on the black hole horizon differ from the Bekenstein-Hawking entropy through terms proportional to the first derivative of the area along the null geodesics. Two different definitions of the dynamical entropy are identified, deduced from gravity symplectic potentials providing a suitable notion of gravitational flux which vanish on non-expanding horizons. The first one is proposed as a definition of the entropy for dynamical black holes by Wald and Zhang, and it satisfies the physical process first law locally. The second one vanishes on any cross section of Minkowski's light cone. I study general properties of its balance law. In particular, I look at first order perturbations around a non expanding horizon. Furthermore, I show that the dynamical entropy increases on the event horizon formed by a spherical symmetric collapse between the two stationary states of vanishing flux, i.e the initial flat light cone and the final stationary black hole. I compare this process to a phase transition, in which the symmetry group of the stationary black hole phase is enlarged by the supertranslations.  
\end{abstract}

\tableofcontents

\section{Introduction}

The covariant phase space formalism developed by Wald and collaborators \cite{lee1990local, crnkovic1987covariant, ashtekar1991covariant, iyer1994some, iyer1997lagrangian,wald2000general, compere2019advanced, harlow2020covariant} has been a powerful tool to study gauge theories with boundaries \cite{iyer1995comparison, donnelly2016local, speranza2018local, harlow2020covariant, freidel2020edge, freidel2020edge2, freidel2021edge, freidel2021extended, chandrasekaran2021general}, in particular black holes thermodynamics \cite{iyer1994some, wald1993black, jacobson1994black, gao2001physical, gao2003first, wald2018kerr}. A key insight from Wald was to understand black hole entropy as a Noether charge of an arbitrary theory of gravity \cite{wald1993black, iyer1994some}. By integrating the symplectic form of the theory contracted with the horizon Killing field on a Cauchy surface between spacelike infinity and the bifurcation surface of the black hole, he was able to relate the variations in phase space of the asymptotic charges, as the ADM mass and angular momentum, to the Noether charge on the bifurcation surface, the entropy (times the Hawking temperature) \cite{wald1993black}, and hence to recover the phase space first law of black hole mechanics \cite{bardeen1973four}. This derivation has been extended in \cite{prabhu2017first} if additional fields with internal (gauge) degrees of freedom are present. However, while the first law can be understood as an equality between the variation of the black hole parameters $M, J, Q$ evaluated at spatial infinity and the area $A$ at the horizon obtained from a general perturbation of the background solution, there exists a second version of this law, known as the physical process first law \cite{hawking1972energy}. It has been derived in \cite{wald1994quantum} and extended to the case where the black hole is charged in \cite{gao2001physical} and \cite{Rignon-Bret:2023lyn}, and states how the black hole entropy is modified when some matter falls into the black hole. The relation between the two versions of the first law is subtle \cite{wald1994quantum, mishra2018physical, sarkar2019black}. Furthermore, it is worth pointing out that other entropy laws have been worked up for dynamical horizons \cite{ashtekar2002dynamical, ashtekar2004isolated}, which are not null but spacelike and foliated by marginally trapped surfaces, and future holographic screens \cite{bousso2015new, bousso2015proof}. The validity of the second law of thermodynamics has also been enlarged to more general theories of gravity \cite{bhattacharjee2016entropy,Bhattacharjee:2015yaa, Chatterjee:2011wj} and investigated for scalar-tensor gravity in \cite{Bhattacharya:2018xlq,Dey:2021rke, Bhattacharya:2022mnb}.

\vspace{0.3 cm}

However, unlike the equilibrium state version of the first law which involves asymptotic charges, the physical process first law is local and derived only from the physics on the event horizon. This local balance law has many interesting features, and led to investigations for further relations between thermodynamics and null hypersurfaces geometry well beyond the range of black holes event horizons. Hence, this work is part of the many attempts of describing some geometric properties of arbitrary null hypersurfaces through thermodynamic relations \cite{chakraborty2015thermodynamical, chakraborty2015gravitational, adami2022null, dey2020covariant}. A more complete approach to these problems is made possible by recent results on the geometry of null hypersurfaces, and the definition of suitable gravitational fluxes and charges on them, particularly on perturbed stationary horizons \cite{Duval:2014uva, Duval:2014lpa, chandrasekaran2018symmetries, Ciambelli:2018wre, Ciambelli:2019lap, Donnay:2019jiz, hopfmuller2017gravity, hopfmuller2018null, Grumiller:2019fmp, oliveri2020boundary, ashtekar2022non, ashtekar2022charges, Odak2022}. These new developments are partially motivated by well known results concerning the relation between gravity and thermodynamics, starting from the laws of black hole thermodynamics. One particularly spectacular result indicating the deep connection between general relativity and thermodynamics is the derivation of the Einstein relation from the Clausius relation by Jacobson \cite{jacobson1995thermodynamics}. Similarly to the physical process first law, this derivation uses the specific form of the Raychaudhuri equation in order to relate the entropy variation, given by the geometry, to the energy variation obtained from the stress energy tensor of the matter crossing the null hypersurface. More generally, it is also known that this equation can be written as a specific instance of a Noether flux balance law \cite{hopfmuller2018null} on an arbitrary null hypersurface $\cN$

\begin{equation}
    dq_\xi \stackrel{\cN}= F_\xi + T_{\mu \nu} \xi^\mu n^\nu \epsilon_\cN
    \label{master equation}
\end{equation}
where $F_\xi$ is the gravitational flux along the diffeomorphism $\xi$ obtained from a well chosen pre-symplectic potential $\Theta$, and $q_\xi$ is the improved Noether charge density \cite{harlow2020covariant, Shi:2020csw, freidel2021edge, freidel2020edge2} associated to the pre-symplectic potential. These improved Noether densities are integrated on a codimension two, surface, the corner. Recently, these charges and their algebra have been treated carefully in the literature, aiming to understand better quantum gravity \cite{donnelly2016local, speranza2018local, Ciambelli:2021nmv,  Ciambelli:2022cfr, Freidel:2023bnj}. For a nice review on the corner proposal, see \cite{Ciambelli:2022vot}. 

\vspace{0.3 cm}

The flux term $F_\xi$ should be written in the canonical form $P \pounds_\xi Q$, where $P$ and $Q$ are canonical pairs and depend only on the intrinsic and extrinsic geometry of the null hypersurface $\cN$. However, while in most work Equation \eqref{master equation} for null future pointing vector $\xi$ is regarded as a first law near equilibrium, we would like to stress that it can also be interpreted as a general balance equation of the entropy written in the common form

\begin{equation}
    dS = S_e + S_c = \frac{\mathcal{Q}}{T_{ext}} + S_c
    \label{second law thermo}
\end{equation}
where $\mathcal{Q}$ is some infinitesimal heat flux flowing into the system and $S_c$ is the infinitesimal entropy creation term, with $S_c \geq 0$. The entropy $S$ is the gravitational charge $q_\xi$, and its variations are given by \eqref{master equation}. It generalizes the idea of identifying the entropy of stationary black holes to the Noether charge associated to null future pointing Killing field on the horizon \cite{wald1993black, iyer1994some, jacobson1994black}. If no matter were present, the entropy variation would be entirely given by the pullback on the boundary $\cN$ of the Noether current constructed out of the gravitational Lagrangian and some pre-symplectic potential. It is analogous to a heat current because it describes the propagation of the microscopical (gravitational) degrees of freedom through the boundary. This flux is deduced from a suitable choice of pre-symplectic potential $\Theta$. Ideally, we would like to disentangle the gauge degrees of freedom from the physical ones, and express $\Theta$ only with the true physical data in order to get a physical flux. Furthermore, we should impose that on any stationary solution, our flux $F_\xi$ vanishes for any boundary generator $\xi$. Hence, a good candidate may be a pre-symplectic potential singled out by the Wald-Zoupas procedure \cite{wald2000general}. Now, if there is matter in play, we should take into account its propagating degrees of freedom too. They do not appear neither in the free gravity Noether charge nor in the gravitational flux $F_\xi$, but they also contribute to the charge variation. In general, irreversibility comes from the presence of degrees of freedom not taken into account into the description of the system (belonging to some environment for instance) which interact with the degrees of freedom of interest. Here, some matter interacts with the gravitational degrees of freedom through the presence of the $T_{\mu \nu} \xi^\mu n^\nu \eps_\cN$ term, analogous to a dissipation term. Indeed, as the null energy conditions are satisfied for generic matter, this term is positive, making relevant to interpret it as an entropy creation term. Hence, the positive energy conditions appear as an essential ingredient to make sense of these entropy balance laws. It is not surprising however, it is well known that the null energy conditions play a key role in the derivations of the area theorem in classical general relativity and for higher curvature theories \cite{Hawking:1971tu, Wall:2015raa, Wall:2011hj}. It also has been derived \cite{Parikh:2015ret} that the null energy conditions, usually associated to the properties of the matter fields, could be derived from assumptions about the validity of the second law for gravity. 

\vspace{0.3 cm}

In section \ref{section2}, we will review the construction of general balance equations for general tensor field theories from the Noether current, focusing on the necessary conditions which must be satisfied in order to write them, and on physical motivations. We usually understand Noether charges as global, with the exception of local gauge symmetries, for which Noether charges are boundary terms. For general theories, we will express the Noether charge variation as a sum of a boundary flux given by the pullback of the Noether current on the boundary and a non-equilibrium term arising if the equations of motion are not satisfied. The Noether charge is conserved if the system is closed and the equations of motion are satisfied. We will give examples for theories described by Lagrangians with a background metric and for theories with covariant Lagrangian. In the latter case, we will explain how this balance law reduces to an entropy law \eqref{second law thermo} if the null energy conditions are satisfied.

\vspace{0.3 cm}

Next, in section \ref{section4}, we will work out different pre-symplectic potentials obtained from the Einstein-Hilbert Lagrangian, leading to different balance laws. However, these pre-symplectic potentials must have the physical meaning of gravitational fluxes, written in the form $P \pounds_\xi Q$, where both $Q$ and $P$ must be covariant with respect to the set of diffeomorphisms which preserves some background structure on the null hypersurface $\cN$. For complete null hypersurfaces at finite distance with topology $B \times \mathbb{R}$ where $B$ is some compact space, these diffeomorphisms are spanned by the superrotations and the supertranslations, the latter being divided into the affine supertranslations and the Weyl supertranslations. Together, they form the BMSW symmetry group \cite{chandrasekaran2018symmetries, Freidel:2021fxf}. Furthermore, we expect that these fluxes vanish in a  stationary spacetime, where no flux is expected. In \cite{chandrasekaran2018symmetries}, the authors applied the Wald-Zoupas procedure to generic null hypersurfaces at finite distance and find an expression for the charge and the flux satisfying the previous requirement, the latter vanishing on non-expanding horizons. we call this flux the Dirichlet flux because it vanishes when the Dirichlet boundary conditions are satisfied. In a $D$-dimensional spacetime, the Dirichlet flux can be written as

\begin{equation}
    F_\xi^D =  \frac{1}{16 \pi}(\sigma^{\mu \nu} - \frac{D - 3}{D - 2} \theta \gamma^{\mu \nu}) \pounds_\xi \gamma_{\mu \nu} \eps_\cN
    \label{fluxDirichlet}
\end{equation}
It can be interpreted as the heat flux flowing through the null hypersurface in \eqref{master equation}, while the improved Noether charge that we get from this pre-symplectic potential 
and associated to the null future pointing diffeomorphism is the entropy. On the event horizon of a black hole perturbed by some incoming matter, the heat flux turns out to be of second order in the stress energy perturbation and so the charge variation only comes from the entropy creation term at first order. In this set-up, if the horizon is affinely parameterized by the coordinate $v$, the diffeomorphism $\xi^\m = \kappa v (\frac{\p}{\p v})$ is null and is a Killing field at first order. The improved Noether charge is the entropy on sections of constant $v$, and turns out to be

\begin{equation}
    S^D = \frac{1}{4} (A - v \frac{dA}{dv})
    \label{Dirichletentropy}
\end{equation}
This entropy formula for dynamical black holes has been priorly proposed by Wald and Zhang from an independent and more general approach \cite{Wald2023}. It has also been studied independently in \cite{Visser2023}. This is a local and dynamical definition of entropy, relevant for a perturbed Killing horizon, i.e if we are close but not exactly at equilibrium. When studying the physical process version of the first law, the term $v \frac{dA}{dv}$ in \eqref{Dirichletentropy} is often disregarded because the charge is integrated up to $v = 0$, close to the bifurcation surface \cite{wald1994quantum}. However, if we decide to keep it in the definition of the entropy, the master equation \eqref{master equation} can be written everywhere on the perturbed Killing horizon at first order as

\begin{equation}
    T_H \Delta S^D = \Delta M - \Omega_H \Delta J - \Phi_H \Delta Q
    \label{PPFL0}
\end{equation}
where $M, J, Q, A$ are respectively the mass, angular momentum, charge and area of the black hole, and $T_H, \Omega_H, \phi_H$ are its Hawking temperature, horizon's angular velocity and horizon's electric potential. However, one inconvenient of the flux formula \eqref{fluxDirichlet} is that it does not vanish anywhere on an spacetime which does not have any non-expanding horizon. In particular, it does not vanish on the simplest null hypersurface embedded in Minkowski spacetime with compact cross sections, which is the light cone. Worse still, this entropy is negative and decreases over successive cross sections of the Minkowski light cone. Hence, we are physically motivated to find a charge which vanishes on such a solution, increases on a general class of future complete null hypersurfaces (if the null energy conditions are satisfied) and gives non vanishing flux only when spacetime is bent and twisted due to incoming fluxes of matter, until to eventually settle down to a black hole. As in thermodynamics, it is sometimes useful to proceed to a Legendre transformation of the symplectic potential in order to get a vanishing flux on a desired dynamical process. In \cite{Odak2022}, the York boundary condition fixes the conformal codimension two metric $\hat{\gamma}_{\mu \nu}$ and the expansion $\theta$ as configuration variables $Q$, in opposition to the Dirichlet boundary conditions treating the whole codimension two metric components as configuration variables. If we proceed this way (and choose the normal to be in the form $n^\m = v \p_v^\m$) we get the following York flux 

\begin{equation}
    F_\xi^Y =  \frac{1}{16 \pi} (\eps_\cN \sigma_n^{\m \n} \pounds_\xi \gamma_{\m \n} + 2 \frac{D-3}{D-2} \eps_\cN \pounds_\xi \theta_n)
    \label{fluxyork}
\end{equation}
for non anomalous diffeomorphisms $\xi$, which form a subset of diffeomorphisms belonging to the BMSW group and preserves the location of the boundary of the null hypersurface. This subset is spanned by the the superrotations and the Weyl supertranslations. The charge generated by the anomalous free Weyl supertranslation null vector $\xi^\mu = \kappa v (\frac{\p}{\p v})^\m$ on cross sections of constant $v$ is the entropy 

\begin{equation}
    S^Y = \frac{1}{4}(A - \frac{1}{D-2}v \frac{dA}{dv})
    \label{chargeyork}
\end{equation}
The flux \eqref{fluxyork} is similar to the one introduced in \cite{hopfmuller2018null}, but here we restricted ourselves to the covariant phase space of \cite{chandrasekaran2018symmetries}, which simplifies and specializes the expressions for the charges and fluxes. The York's flux \eqref{fluxyork} and the York's charges, including the entropy, vanish on Minkowski's outgoing light cone, while Dirichlet's flux \eqref{fluxDirichlet} and Dirichlet's charge \eqref{Dirichletentropy} do not. This is a desired property, as we do not expect any gravitational flux or gravitational charge in Minkowski's spacetime. Of course, the York's flux \eqref{fluxyork} also vanishes on non-expanding horizons. Furthermore, we will study in detail this flux and identify the cases for which it is positive or null. In particular, on a null hypersurface $\cN$ with topology $B \times \mathbb{R}$ that is future complete, we prove that the variation of the York charges associated to future pointing Weyl supertranslations on the cross sections of $\cN$ are always positive if the null energy conditions hold. In addition, we prove that the York dynamical entropy always increases during a spherically symmetric collapse up to the formation of a black hole \eqref{chargeyork}, in which case the value of the charge on the stationary horizon is $\frac{A}{4}$. In particular, the York entropy \eqref{chargeyork} of the black hole horizon does not increase and remains null as long as matter has not crossed the event horizon. The dynamical geometric parameter $\theta_n$ evolves from $D - 2$ on the initial Minkowski light cone to $\theta_n = 0$ on the late stationary horizon. Hence, we argue that the formation of a spherically symmetric black hole might be understood as a phase transition between two stationary states, with order parameter given by the expansion. The stationary black hole is the phase of high symmetry, and the symmetry group preserving the pullback of the metric on the event horizon is $SO(D-1) \ltimes \mathbb{R}_W^S \ltimes \mathbb{R}_T^S$ \cite{chandrasekaran2018symmetries, Freidel:2021fxf}, while the flat light cone is the low symmetry phase for which symmetry group $SO(D-1) \ltimes \mathbb{R}_W^S \ltimes \mathbb{R}_T^S$ is broken and all the supertranslations are eliminated.

\vspace{0.3 cm}

We also analyze master equation \eqref{master equation} for the York boundary conditions at first order in perturbation around a non expanding horizon. Unlike the Dirichlet case, the charge variation between two cross sections of constant $v$ is not entirely given by the matter term at first order. Indeed, there is an additional term taking into account the York flux \eqref{fluxyork} at first order. On a portion of the horizon without matter, near equilibrium, we can write at first order 

\be
    T_H \Delta S^Y = F_\xi = \mathcal{Q}^Y = \Delta U_{grav}^Y
    \label{internalenergyhawk}
\ee
where $U_{grav}^Y = \frac{1}{4} k_B T_H \frac{D - 3}{D - 2} A$ and $\mathcal{Q}^Y$ is the heat flux. This law is analogous to a first law of thermodynamics (in vacuum), where the non vanishing gravitational flux \eqref{fluxyork} contributes to increase a local quantity $U_{grav}^Y$ that can be interpreted of an internal energy associated to the gravitational degrees of freedom on the dynamical event horizon. 

\vspace{0.3 cm}

Except in some rare occasions where we will restore the fundamental constants, we will assume in the rest of the manuscript that $G = c = \hbar = k_B = 1$. 

\section{Second law from Noether charge analysis}
\label{section2}

\subsection {The Noether current}

In this section, we derive local balance laws of surface charges for general field theories in the presence of background fields. We obtain general conservation equations and relations analogous to Bianchi identities in gauge theories. Most of these results and methods are well known, but the emphasis is put on the presence of general background fields. The main point of this section is the interpretation of these balance equations focusing on the role played by the surface charge and the different pieces contributing to its variation. We first need to study the different symmetries of a theory with Lagrangian $L$ and the structure of the resulting Noether current. Let us assume that our theory describes some dynamical fields $\phi$ that are part of our configuration space propagating next to fixed background fields $\chi$ on a manifold $\mathcal{M}$ \footnote{We can consider several fields $\phi^i$ and $\chi^j$, and from now on, the sum over all the different fields in the following equations will be implicit, we will not mention the indices $i$ and $j$ anymore.} with local volume form $\eps_\mathcal{M}$. The Lagrangian $L(\phi, \chi)$ describing our theory is written only in terms of these fields, and is an analytic function of them and their derivatives. By varying the action we obtain the well known identity
\be
    \d L = \frac{\d L}{\d \phi} \d \phi + d \Theta
    \label{varprinc}
\ee
where $\Theta$ is some pre-symplectic potential. From now, to simplify the writing, we will refer to $\Theta$ as a symplectic potential rather than a pre-symplectic potential. Furthermore, if we contract \eqref{varprinc} with some diffeomorphism $\xi$ we get
\begin{equation}
    \int_\mathcal{M} d(i_\xi L - I_\xi \Theta) = \int_\mathcal{M} \frac{\delta L}{\delta \phi} \cdot \pounds_\xi \phi + \frac{\p L}{\p \chi} \cdot \pounds_\xi \chi
    \label{dj}
\end{equation}
where $I_\xi$ is the field space interior product associated to the configuration space vector $X_\xi$, defined by

\be
    X_\xi = \int_\mathcal{M} d^D x \pounds_\xi \phi \frac{\d}{\d \phi}
    \label{defxxi}
\ee
If all the fields in \eqref{dj} are dynamical fields, such that there is no field $\chi$, all the diffeomorphisms are symmetries of our theory, as $\d_\xi L = \pounds_\xi L = d i_\xi L$ is a boundary term. However, it can happen that the total Lagrangian $L$ of all the physical fields involved is unknown, or that the equations of motion of some fields are too complicated to solve, such that we prefer not to use their equations of motions and fixing $\chi$ a priori. The example on which we will mainly focus in the following is the case where the background field $\chi$ is the metric $g$. In any case, we have to add the contribution of the diffeormophism acting on these background fields to the field space Lie derivative $\d_\xi$ in order to get the usual spacetime Lie derivative. Therefore, we get : $\d_\xi L = \pounds_\xi L - \frac{\p L}{\p \chi} \pounds_\xi \chi$ that is not a boundary term anymore in general \footnote{In fact, if the anomaly is a boundary term, i.e if $\Delta_\xi L = da_\xi$ for all $\xi$, then $\d_\xi L$ is a boundary term and $a_\xi$ should be included in the definition of the Noether current if we want it to be conserved \cite{Freidel:2021fxf, Odak:2022ndm}. It appears if the source of anomaly is the background structure that we introduce to define the boundary.}, so not all the diffeomorphisms are symmetries of our theory \footnote{Here we should underline that we used the partial derivative and not the functional derivative. In other words, each background field $\chi$ and its covariant derivatives are treated as independent field and so the implicit sum on the tensor fields takes into account the successive covariant derivatives of each background $\chi$.}. The term $\frac{\p L}{\p \chi} \cdot \pounds_\xi \chi$ is called an anomaly, and it can prevent some diffeomorphisms $\xi$ from being a symmetry. We can define the anomaly operator acting on tensors as $\Delta_\xi = \delta_\xi - \pounds_\xi$ \cite{hopfmuller2018null}. However, we can still have $\frac{\p L}{\p \chi} \cdot \pounds_\xi \chi = 0$ if $\pounds_\xi \chi = 0$, i.e if a subclass of diffeomorphisms leaves the environment $\chi$ invariant. In the case where $\chi$ is the background metric, the diffeomorphisms satisfying such a condition are the Killing fields. These diffeomorphisms preserving the background structure are symmetries of our theory, and then the Noether current

\begin{equation}
    j_\xi = I_\xi \Theta - i_\xi L
    \label{Noethercurrent}
\end{equation}
is conserved on-shell \eqref{dj}. From this Noether current, we now review the general analysis leading to Bianchi identities and balance laws, but we take great care of the terms containing the information about the background structure. In general, the Lie derivative of any tensor field can be expressed as a sum of terms proportional to $\xi$ and to first derivatives of $\xi$. Hence, for general tensor fields $\phi$ and $\chi$ we have\footnote{These notations are informal, but keeping all the indices at the right place would make it harder to follow. More precisely, if $\phi = \phi_{\m_1 \m_2 \cdots \m_n}$ is a $n$-covariant tensor, its Lie derivative $\pounds_\xi \phi_{\m_1 \m_2 \cdots \m_n} = \xi^\alpha \nabla_\alpha \phi_{\m_1 \m_2 \cdots \m_n} + \phi_{\alpha \m_2 \cdots \m_n} \nabla_{\m_1} \xi^\alpha + \phi_{\m_1 \alpha \cdots \m_n} \nabla_{\m_2} \xi^\alpha + \cdots = \xi^\alpha \nabla_\alpha \phi_{\m_1 \m_2 \cdots \m_n} + \sum_{\alpha = \m_1}^{\alpha = \m_n} \phi_{\m_1 \cdots \nu \cdots \m_n} \nabla_\alpha \xi^\nu$, and so the Lie derivative is a finite sum of terms proportional to $\xi^\mu$ and $\nabla_\nu \xi^\mu$. It is straightforward to verify it is also true for a generic $p$-covariant and $q$-contravariant tensor, and can be written as $\pounds_\xi \phi =  [\phi] \cdot \nabla \xi + \nabla \phi \cdot \xi $ where the fields $[\phi]$ are the set of coefficients in front of the $\nabla_\mu \xi^\nu$ terms.
\label{footnote}
}

\begin{align}
    \pounds_\xi \phi &= [\phi] \cdot \nabla \xi + \nabla \phi \cdot \xi \nn \\
    \pounds_\xi \chi &= [\chi] \cdot \nabla \xi + \nabla \chi \cdot \xi 
\end{align}
where $[\phi]$ and $[\chi]$ are coefficients in front of the $\nabla \xi$ terms (basically, they are sums of $\phi$ and $\chi$ where each term is contracted with $\nabla \xi$ through different indices, as explained in the footnote \ref{footnote}). Then, we can integrate by part the terms $[\phi] \cdot \nabla \xi$ and $[\chi] \cdot \nabla \xi$ and get a sum of a boundary term and a term linear in $\xi$. Thus we can write \ref{dj} as

\begin{equation}
    \int_\mathcal{M} d(i_\xi L - I_\xi \Theta -i_{K_\xi} \epsilon_\mathcal{M}) = \int_\mathcal{M} \epsilon_\mathcal{M} \Bigl[ - \nabla \cdot (\frac{\delta \mathcal{L}}{\delta \phi} \cdot [\phi] + \frac{\p \mathcal{L}}{\p \chi} \cdot [\chi]) + \frac{\delta \mathcal{L}}{\delta \phi} \cdot \nabla \phi + \frac{\p \mathcal{L}}{\p \chi} \cdot \nabla \chi \Bigr] \cdot \xi
    \label{bulkboundaryeq}
\end{equation}
where the Lagrangian density $\mathcal{L}$ is defined through $L = \mathcal{L} \eps_\mathcal{M}$ and the vector $K_\xi$ is given by

\begin{equation}
    K_\xi = \frac{\d \mathcal{L}}{\d \phi} \cdot [\phi] \cdot \xi + \frac{\p \mathcal{L}}{\p \chi} \cdot [\chi] \cdot \xi
    \label{K}
\end{equation}
The left hand side of \ref{bulkboundaryeq} is a boundary term while the right hand side is a bulk term. We can vary arbitrarily $\xi$ in the bulk while keeping it constant on the boundary $\partial M$. Therefore, the only way for the equality to hold is to make both integrands of \ref{bulkboundaryeq} vanish. Hence we obtain the two following relations

\begin{align}
    0 &= - \nabla \cdot (\frac{\delta \mathcal{L}}{\delta \phi} \cdot [\phi] + \frac{\p \mathcal{L}}{\p \chi} \cdot [\chi]) + \frac{\delta \mathcal{L}}{\delta \phi} \cdot \nabla \phi + \frac{\p \mathcal{L}}{\p \chi} \cdot \nabla \chi \nn \\
    j_\xi &= I_\xi \Theta - i_\xi L = - (\frac{\d \mathcal{L}}{\d \phi} \cdot [\phi] \cdot \xi + \frac{\p \mathcal{L}}{\p \chi} \cdot [\chi] \cdot \xi) \cdot \eps_\mathcal{M} + dq_\xi = - K_\xi \cdot \eps_\mathcal{M}+ dq_\xi
    \label{Noetherid}
\end{align}
The first equality is similar to the relation between the equations of motions that we get in Noether's second theorem, so we will refer to it as a Bianchi identity. One illustrative example of this Bianchi identity \eqref{Noetherid} in the presence of background fields is to take the metric for $\chi$, as we should do for any Lagrangian describing the dynamic of some matter field $\phi$ without taking into account the back reaction of the metric. In this case, the first equation of \eqref{Noetherid} simply tells us that the stress energy tensor is conserved on shell. From a theorem due to Wald \cite{wald1990identically}, the relation $d(i_\xi L - I_\xi \Theta -i_{K_\xi} \epsilon_M) = 0$ implies that $i_\xi L - I_\xi \Theta -i_{K_\xi} \epsilon_M = dq_\xi$ with $q_\xi$ being constructed from the fields $\phi$ and $\chi$ and their derivatives, and justifies the second relation of \eqref{Noetherid}. This second identity is similar to some expressions for currents obtained in \cite{iyer1994some, iyer1997lagrangian}, where here the dynamical and background fields are supposed to be very general. From \eqref{Noetherid}, we read that the current is not a boundary term either if the equations of motion are not satisfied or if there are some background fields in the description of our theory. In both cases, it means that there exist some degrees of freedom which were not taken into account in the description of our system, either because we missed a piece in the Lagrangian or because some fields in the Lagrangian we used were not dynamical. 

\vspace{0.3 cm}

We can now use this Noether current $j_\xi$ to write balance laws for Noether charges associated with $\xi$ with or without the presence of background fields $\chi$. First, we look at theories where some fields $\chi$ are present, but then $\xi$ must be a symmetry of the background, i.e $\pounds_\xi \chi = 0$. Thus, we get from \eqref{dj}
\be
    dj_\xi = - \frac{\d L}{\d \phi} \pounds_\xi \phi
    \label{dj2}
\ee
Then, we integrate \eqref{dj2} over a manifold $\mathcal{M}$ whose boundary is composed of the initial and final spacelike slices $\Sigma_1$ and $\Sigma_2$ and a timelike or null boundary $\cN$ joining the spacelike slices. Thus, we have 
\be
    \Delta Q_\xi = Q_{\xi}^{\Sigma_2} - Q_{\xi}^{\Sigma_1} = \int_\cN j_\xi + \int_{\mathcal{M}} \frac{\d L}{\d \phi} \pounds_\xi \phi
    \label{balancelawback}
\ee
where we defined 
\be
    Q_\xi^{\Sigma_i} = -\int_{\Sigma_i} j_\xi 
\ee
as the Noether charge evaluated on the spacelike surface $\Sigma_i$, $i = 1,2$. If the system is closed, which means that the pullback of the Noether current $j_\xi$ on $\cN$ vanishes, then the charge $Q_\Sigma$ is conserved on-shell. Furthermore, if $\xi$ is tangent to $\cN$, \eqref{balancelawback} can be written as 
\be
    \Delta Q_\xi = Q_{\xi}^{\Sigma_2} - Q_{\xi}^{\Sigma_1} = \int_\cN I_\xi \Theta + \int_{\mathcal{M}} \frac{\d L}{\d \phi} \pounds_\xi \phi
    \label{balancelawback2}
\ee
In any case, we see that \eqref{balancelawback} and \eqref{balancelawback2} have both the same structure as the standard balance formula
\begin{equation}
     \Delta A = A_{ex} + A_c
     \label{balanceeq}
\end{equation}
with $A$ being the charge $Q_\xi^{\Sigma_i}$, the integral on $\cN$ of the Noether current $j_\xi$ is the flux of the charge $Q_\xi$ through the boundaries of the system, that we denote $A_{ex}$, and the integral on $\mathcal{M}$ of $\frac{\d L}{\d \phi} \pounds_\xi \phi$ is the creation term $A_c$. The last term arises because the equation of motion are not satisfied everywhere in $\mathcal{M}$, i.e the fields $\phi$ are coupled to other fields that were not present in the initial Lagrangian.  The fact that the equations of motion are not satisfied implies that the action from which we started is not stationary, and we can therefore consider the term $\frac{\d L}{\d \phi} \pounds_\xi \phi$ as an out-of-equilibrium term. On-shell, this out-of-equilibrium term vanishes and the charge variation
\be
    \Delta Q_\xi = \int_{\Sigma_2} \frac{\p \mathcal{L}}{\p \chi} \cdot [\chi] \cdot \xi - dq_\xi - \int_{\Sigma_1} \frac{\p \mathcal{L}}{\p \chi} \cdot [\chi] \cdot \xi - dq_\xi = \int_{\cN} I_\xi \Theta
    \label{pullbackback3}
\ee
is simply given by the flux of the Noether current on the boundary $\cN$.  Furthermore, if there were no background structure $\chi$, as in a diffeomorphism invariant theory like general relativity, the charge would be given only by a boundary term, as expected.   
A simple example for a theory in the presence of a background structure is given by the balance law of electromagnetic energy. The dynamical field is the potential $A_\m$ while the background field is the metric $g_{\m \n}$. In this case, the charge is the electromagnetic energy $e$, whose variation is given by the flux of the Pointing vector across the boundary $\cN$ of normal $n$ and a creation term indicating the energy transfer from the charge current $J^i$ to the electric field $E_i$, appearing because we did not take into account the charged matter degrees of freedom when we worked out the Noether current from the Lagrangian of the free electromagnetic field. A detailed calculation is presented in Appendix.\ref{AppA}. The result is the well-known balance equation for electromagnetic energy 
\be
    \Delta e = - \int_{\cN} \pi_i n^i \eps_{\cN} + \int_{\mathcal{M}} J^i E_i \eps_{\mathcal{M}}
\ee
Here the variation of electromagnetic energy $A$ is the sum of a boundary flux $A_{ex}$ given by the integral of the Poynting vector on the boundary $\cN$ and a creation term indicating the energy transfer from the charges to the field, necessary in order to take into account the matter degrees of freedom interacting with the electromagnetic field through the electric charge current $\Vec{J}$. 

\vspace{0.3 cm}

For a diffeomorphism invariant theory like general relativity, \eqref{dj2} holds for any diffeomorphism $\xi$ because there is no background structure in the bulk Lagrangian. The only contribution of the background fields comes from the non equilibrium condition $\frac{\d L}{\d \phi} \neq 0$, reflecting that we missed the description of some fields in our initial Lagrangian. In that case, \eqref{balancelawback2} reduces to a balance law on the boundary $\cN$, because the Noether charge is now a boundary term and so the integrals on the spacelike surfaces $\Sigma_i$ reduce to an integral on the corners. We can call $S_i$ the intersection between the spacelike slice $\Sigma_i$ and the null or timelike  boundary $\cN$. For a diffeomorphism invariant theory, the metric is part of the dynamical fields $\phi$. The non equilibrium term $\frac{\d L}{\d \phi} \pounds_\xi \phi$ can be expressed as a boundary term on $\cN$ thanks to the general Bianchi identities \eqref{Noetherid}, giving 
\be
   \frac{\d L}{\d \phi} \pounds_\xi \phi = d [(\frac{\d \mathcal{L}}{\d \phi} \cdot [\phi] \cdot \xi) \cdot \eps_\mathcal{M}]
\ee
Here and in the remaining of the paper, we will restrict ourselves to $\phi = g$, and so the only dynamical field is the metric. Therefore, the non-equilibrium term $\frac{\d L}{\d \phi} \pounds_\xi \phi$ is expressed as a flux of stress energy across the boundary. Then the balance equation \eqref{balancelawback2} reduces to the pullback of \eqref{Noetherid} on an hypersurface $\cal N$ of normal $n$, on which $\xi$ is tangent \footnote{Here we chose to define $\eps_\cN$ as $\eps_\mathcal{M} = - n \wedge \eps_\cN$ because we will work with null hypersurfaces in the following, and we will associate to the normal $n$ an auxiliary null vector $l$ such that $n \cdot l = -1$. If the hypersurface were timelike or spacelike, we would have chosen instead $\eps_\mathcal{M} = n \wedge \eps_\cN$.}. The charge variation is then given by 
\begin{equation}
    \int_{S_2} q_\xi - \int_{S_1} q_\xi = \int_{\cN} dq_\xi = \int_{\cN} I_\xi \Theta - \eps_\cN K_\xi \cdot n = \int_{\cN} I_\xi \Theta + \int_{\cN} T_{\m \n} \xi^\m n^\n \eps_{\cN}
    \label{pullbackcurrent}
\end{equation}
where $\eps_\cN$ is a volume form on the hypersurface of normal $n$, defined through the relation $\eps_\mathcal{M} = - n \wedge \eps_\cN$. One of the most interesting case is to take a null boundary $\cN$. If $\xi$ is a future pointing infinitesimal diffeomorphism tangent to $\cN$ \footnote{And so $\xi$ must also be null.} and if the null energy conditions are satisfied, then the creation term $A_c = T_{\m \n} \xi^\n n^\n \eps_\cN$ is positive \footnote{It may be interesting to notice that we can also take the pullback of the Noether current on $\cN$ if background fields $\chi$ are present, in order to get a balance law for the corner term $q_\xi$. The integral of the corner term $q_\xi$ on a codimension two surface is not the Noether charge anymore. However, this balance law is similar to the one we get for diffeomorphism invariant theories \eqref{pullbackcurrent} with an additional term in $K_\xi$ coming from the presence of this background field $\chi$.}. Therefore, the equations \eqref{balancelawback2} or \eqref{pullbackcurrent} look exactly like an entropy balance equation 
\be
    \Delta S = S_e + S_c
\ee
with $S_c \geq 0$. The similarity between the balance law for a Noether charge associated to a null future pointing diffeomorphism and the usual entropy balance law makes it a natural candidate for entropy in classical gravitational theories, provided that the null energy conditions are satisfied. On the particular case of a Killing horizon, it has been well known that the gravitational Noether charge associated to the future pointing Killing field on the horizon is the entropy of a stationary black hole \cite{wald1993black, iyer1994some}.

\subsection{Improved Noether charge and choice of polarization}

We remind in this short subsection some results about the Noether charge \cite{noether1971invariant} and its symplectic flux for a diffeomorphism invariant theory. These Noether charges are obtained from the Lagrangian $L$ and symplectic potential $\Theta$ of our theory. However, it is well known \cite{noether1971invariant, lee1990local,jacobson1994black,harlow2020covariant,freidel2020edge, freidel2020edge2,freidel2021extended, chandrasekaran2021general} that the Noether charges are as ambiguous as the Lagrangian $L$ and the symplectic potential $\Theta$ are. First, if we start from a Lagrangian $L$, the symplectic potential defined through \eqref{varprinc} is ambiguous to the addition of an exact form $\Theta \longrightarrow \Theta - d \vartheta$. Second, it is always possible to shift this Lagrangian by an exact form in spacetime $L \longrightarrow L + dl$. This process does not change the equations of motion either, but shifts the symplectic potential by an exact form in field space $\Theta \longrightarrow \Theta + \d l$. If the Lagrangian $L$ is covariant, i.e $\Delta_\xi L = 0$ for any diffeomorphism $\xi$ (and it will be the class of Lagrangian we are considering in the following), we can choose a symplectic potential $\Theta^0$ to get \eqref{Noetherid} evaluated on-shell
\be
    dq_\xi^0 = I_\xi \Theta^0 - i_\xi L
    \label{Noetherchargedef}
\ee
The charge density $q_\xi^0$ \footnote{Sometimes, we refer to the quantities $q_\xi$ as the charge rather than the charge density. Properly speaking, the charge should be given by the integral of the charge density on a manifold.} is also ambiguous up to the addition of an exact form $q_\xi^0 \longrightarrow q_\xi^0 + d Y_\xi$ but this ambiguity is irrelevant for the charge obtained from the integral of $q_\xi^0$ on codimension two compact cross sections. Hence, in general, these ambiguities allow as to shift the symplectic potential $\Theta^0$ by
\be
    \Theta = \Theta^0 + \d l - d \vartheta
    \label{generalsymppot}
\ee
and the charge density obtained from \eqref{Noetherchargedef} and \eqref{generalsymppot} is 
\be
    q_\xi = q_\xi^0 + i_\xi l - I_\xi \vartheta
    \label{generalcharged}
\ee
This is the improved Noether charge formula. The existence of an ambiguity in the definition of the charge is related to the ambiguity in the polarization of our phase space. Indeed, different choices of boundary Lagrangians $l$ and corner terms $\vartheta$ select different configuration and momentum fields in the phase space. We would like to put the symplectic potential in the form $\Theta = P \d Q$, where $P$ and $Q$ can be expressed in terms of the dynamical fields $\phi$ and the background structure. There is not symplectic flux if the boundary conditions are imposed, i.e if $\d Q = 0$ for some chosen phase space polarization. In this case, the system is closed, and if the symplectic potential $\Theta = P \d Q$ is covariant \footnote{It means that the difference between the phase space Lie derivative and the spacetime Lie derivative vanishes, namely $(\d_\xi - \pounds_\xi) \Theta = 0$. In the following, we will restrict ourselves to symplectic potentials satisfying this property.}, then the charge $dq_\xi$ is a Hamiltonian as 
\be
    -I_\xi \omega = \d d q_\xi
\ee
on-shell where $\omega = \d \Theta$ is the symplectic two form obtained from the  potential $\Theta$. In general, however, we have that 
\be
    -I_\xi \omega = d (\d q_\xi - i_\xi \Theta)
\ee
that can be obtained from \eqref{Noethercurrent} by taking the field space variation of \footnote{In this paper, we always assume that $\d \xi = 0$. For generalizations that take into account field dependent diffeomorphisms and anomalies, see \cite{Odak:2022ndm}} and the Noether charge $q_\xi$ is not Hamiltonian. 
If $\xi$ is tangent to the boundary $\cN$, then we can take the pullback of \eqref{Noethercurrent} on $\cN$ and get $dq_\xi \stackrel{\cN}= P \d_\xi Q$. If $Q$ is covariant, then the flux of the charge is just given by
\be
    dq_\xi = P \pounds_\xi Q
    \label{fluxchargesymp} 
\ee
and should vanish for some stationary solution. By stationary solution, we mean a spacetime solution that does not carry any radiation, like a stationary black hole spacetime or flat spacetime. This criterion can help us to choose one polarization over another. The formula \eqref{fluxchargesymp} indicates that the flux vanishes on the boundary either if $P = 0$ or $\pounds_\xi Q = 0$ for any diffeormorphism $\xi$ preserving the background structure of the boundary $\cN$. In the following, we are facing these two cases. 

\section{Dynamical entropies}
\label{section4}

\subsection{Geometry of null hypersurfaces}
\label{section41}
\subsubsection{Definition of the geometric quantities }
\label{section411}

Before getting to the heart of the matter, we should review some basics about the geometry of null hypersurfaces \footnote{See for instance \cite{Odak2022} for more details about the geometry on null hypersurfaces}. Let us consider some pseudo Riemannian manifold $\mathcal{M}$ equipped with a volume form $\epsilon_\mathcal{M}$. Consider also a null "boundary" $\cN$ of $\mathcal{M}$, and choose a future directed null normal $n$ of $\cN$. Then we can define a volume form $\eps_\cN$ on $\cN$ through the relation

\be
    \eps_\mathcal{M} = - n \wedge \eps_\cN
    \label{volumeform}
\ee
We define an auxiliary null vector $l$ such that

\be
    l^\m n_\m \stackrel{\cN}= -1
    \label{normalizationl}
\ee
and the minus sign comes from the fact that we want the vector $l$, as the vector $n$, to be future directed \footnote{More precisely, from \eqref{volumeform} and \eqref{normalizationl}, we can define $\eps_\cN = \pbi{i_l \eps_\mathcal{M}}$. Indeed, taking the pullback on $\cN$ is necessary in order to define in a unique way $\eps_{\cN}$, and because of that the definition of $\eps_\cN$ through \eqref{volumeform} is ambiguous.}. We can complete the basis $(n,l)$ by spacelike vectors $e_A$ tangent to $\cN$, and in addition, we can choose them so that they are orthonormal. On $\cN$, we can define a codimension two form from $n$ and $\eps_\cN$

\be
    \eps_S = i_n \eps_\cN
    \label{epsS}
\ee
or equivalently

\be
    \eps_\cN = -l \wedge \eps_S
    \label{epsNS}
\ee
and we define the expansion $\theta_n$ associated to the normal $n$ on $\cN$ as

\be
    d\eps_S = \theta_n \eps_\cN
    \label{defexpansion}
\ee
The projector on the tangent subspace $Span (e_A)_{A \in (1, \cdots, D - 2)}$ orthogonal to $n$ and $l$ is

\be
    \g_{\m \n} = g_{\m\n} + n_\m l_\n + n_\n l_\m
    \label{projector}
\ee
from which we define the physical quantities on $\cN$ which encapsulate the intrinsic and extrinsic geometry of $\cN$. One of the most important object to achieve this is the Weingarten map
\be
    W_{\m}^\n = \pbi{\nabla_\m} n^\n
\ee
from which all the following relevant quantities can be derived 
\begin{align}
    \g_\m^\alpha \g_\n^\beta \nabla_\m n_\n &= (\sigma_n)_{\m \n} + \frac{1}{D-2} \theta_n \g_{\m \n} \nn \\
    k_n &= - l^\m n^\n \nabla_\n n_\m \nn \\
    \eta_\mu &= \gamma_\mu^\rho l^\sigma \nabla_\rho n_\sigma 
    \label{dictionary}
\end{align}
where $(\sigma_n)_{\m \n}$ is the traceless shear \footnote{From the first equation of \eqref{dictionary}, we can deduce that the expansion $\theta$ is also given by $\theta = \g^{\m \n} \nabla_\m n_\n$. To see the relation with the definition \eqref{defexpansion}, compute on one hand $\pounds_n \eps_{\mathcal{M}} = \nabla_\m n^\m \eps_{\mathcal{M}}$ and on the other hand $\pounds_\n \eps_{\mathcal{M}} = - \pounds_n n \wedge \eps_\cN - n \wedge \pounds_\n \eps_\cN = \omega_n \eps_{\mathcal{M}} + \theta \eps_{\mathcal{M}} - n \wedge i_n d \eps_n$, because $\pounds_\n \eps_\cN = (d i_n + i_n d) \eps_\cN = \theta \eps_\cN + i_n d \eps_N$ and $\omega_n$ is defined as $\pounds_\n n = \omega_n n$. Furthermore, $n \wedge i_n d \eps_\cN = 0$ vanishes because $i_n n = 0$. Hence we understand that $\theta = \nabla_\m n^\m - \omega_n$. But $\nabla_\m n^\m = g^{\m \n} \nabla_\m n_\n$ and we can compute that $\omega_n = - 2 n^{(\m}l^{\n)}\nabla_\m n_\n$, hence we conclude that we also have $\theta = \g^{\m \n} \nabla_\m n_\n$ from \eqref{projector}.} associated to the normal $n$, $k_n$ the unaffinity of $n$ and $\eta_\m$ is the twist.

\subsubsection{Topology}
\label{section412}

In the following, we will be interested in null hypersurfaces (or portion of null hypersurfaces) with topology $B \times \mathbb{R}$, where $B$ is some compact base space \footnote{The compactness is needed in order to get finite charges on the spacelike cross sections.} (that in most applications will be the $D-2$ sphere $S^{D-2}$) as in \cite{chandrasekaran2018symmetries}. Basically, it means that the null hypersurface is foliated by a geodesic congruence such that the geodesics do not cross \footnote{This is why the supertranslations are general symmetries of these hypersurfaces, as we will see in the following.}. By making such a choice, we avoid caustics, and the null geodesics cross any spacelike cross section of $\cN$ only once. Hence, the mathematical objects presented in the section above describing the intrinsic and extrinsic geometry are well-behaved. However, at some point in this paper we will also study null hypersurfaces where the geodesics cross at one point (but has topology $B \times \mathbb{R}$ otherwise). Each geodesic can be equipped by an affine parameter $v$ (there are many choices of affine parameter), such that the null vector $n$ tangent to the geodesics is 

\be
    n = f \frac{d}{dv}
    \label{normalexp}
\ee
with $f \geq 0$ in order to be future directed. If we can extend the affine parameter $v$ for each geodesic to infinity in both directions, then the geodesic congruence of the null hypersurface is said to be complete. We notice that this condition do not depend on our particular choice of affine parameter. If we can extend the affine parameter only to future (past) infinity, then we say that it is future (past) complete. \footnote{For instance, we will study further the outgoing light cone, which is only future complete, i.e we can extend the null geodesics to future infinity but not to past infinity because of the light cone's tip.}. We can study portion of hypersurfaces of topology $B \times \mathbb{R}$, but in this case the geodesic congruence will not be complete. 

\subsubsection{Coordinate choice}

We can then describe any generic metric in the neighbourhood of our null hypersurface $\cN$ located at $u = 0$ using a judicious coordinate system. Even if only covariant quantities are meant to be physical, it is sometimes insightful to work in a well adapted gauge to gain some physical insight. There are many suitable gauge choices, but we would like that the affine parameter $v$ be part of the coordinates. We can choose the Newman-Unti gauge as in \cite{ashtekar2022charges}, because this gauge is conserved by the set of diffeomorphisms preserving the background structure that we will introduction in section.\ref{section32}, see Appendix.\ref{AppB} for further details. If the geodesic congruence of null hypersurface $\cN$ is complete, we can extend the value of the affine parameter in both directions to infinity, and so the neighborhood of the whole hypersurface $\cN$ can be described by the chart $(u, v, x^A)$, and on the null hypersurface $\cN$ located at $u = 0$ with the coordinates $(v, x^A)$, where the $x^A$ are the coordinates describing the spacelike cross sections at constant $v$. Therefore, a generic metric in some small neighborhood of $\cN$ can be written in the Newman-Unti gauge as \footnote{We see in \eqref{NUmetric0} that $g_{vv} = O(u^2)$. This is needed because $v$ is an affine parameter. Indeed, a short calculation shows that $\frac{\p}{\p v}^\n \nabla_\n \frac{\p}{\p v}^\m = \Gamma^{\m}_{vv} = 0$ only if $\p_u g_{vv} = 0$ at $\cN$.}

\be
    g_{\m \n} dx^\m dx^\n = - u^2 F dv^2 - 2 du dv + 2 u P_A dx^A dv + g_{AB} dx^A dx^B
    \label{NUmetric0}
\ee
where $F$, $P_A$ and $g_{AB}$ are $\frac{D(D-1)}{2}$ functions of $v$ and $x^A$ \footnote{As the constraints have not been imposed yet, these $\frac{D(D-1)}{2}$ functions are not independent of each other.}. So, the normal vector tangent to the null geodesics is $n^\m = f \p_v^\m$ and we can define an auxiliary vector $l^\m = \frac{1}{f} \p_u^\m$ and a coordinate basis $\frac{\p}{\p x^A}$ tangent to the constant $v$ cross sections of $\cN$. On $\cN$, located at $u = 0$, the $D$-dimensional metric \eqref{NUmetric0} becomes 

\be
    g_{\m \n} dx^\m dx^\n \vert_{u = 0} = - 2 du dv + g_{AB} dx^A dx^B
    \label{NUmetriconN0}
\ee
Hence, on $\cN$, we have $n_\m = - f \p_\m u$ and $l_\m = - \frac{1}{f} \p_\m v$. We should notice that the fact that $g_{vv} = O(u^2)$ in \eqref{NUmetric0} implies $l^\m \p_\m n^2 = 0$ on $\cN$, and so we have the simpler expression for the unaffinity $k_n$ \eqref{dictionary} 

\be
    k_n = n^\m \p_\m \ln{f}
    \label{unaffinity}
\ee
and so we can check that the coordinate $v$ is affine as the unaffinity of the normal $n^\m = \p_v^\m$ vanishes.

\subsection{Local entropy balance law for Dirichlet flux}
\label{section32}
\subsubsection{Boundary structure and Dirichlet flux and charges}

From now on, and except at the end of this section where we will come back to the more general $D$-dimensional case, we will assume $D = 4$. If we vary the Einstein-Hilbert action on a manifold $\mathcal{M}$ with a null boundary $\cN$ \cite{parattu2016boundary, lehner2016gravitational, oliveri2020boundary}, we get in addition to the equations of motion the exterior derivative of the bare Einstein-Hilbert \footnote{It is the usual boundary term $n^\alpha g^{\m \n} (\nabla_\m g_{\n \alpha} - \nabla_\alpha \delta g_{\m \n})$ that we get from varying $g_{\m \n}$.} symplectic potential integrated on the null boundary $\cN$ \footnote{We can notice that the easiest way to get the symplectic potential on a null hypersurface is to compute first the symplectic potential for tetrad gravity rather than metric gravity, as it is done in \cite{oliveri2020boundary}. However, we should keep in mind that the two symplectic potentials (metric and tetrad) differ by an exact 3-form \cite{DePaoli:2018erh, Oliveri:2020xls}}

\be
    \int_{\cN} \pbi{\Theta}^{EH} = \frac{1}{16 \pi} \int_\cN (\sigma_n^{\m \n} - \frac{1}{2} (2k +\theta_n) \g^{\m \n}) \d \g_{\m \n} \eps_{\cN} + 2 (\eta_\m - \theta_n l_\m) \delta n^\m \eps_\cN + 2 \d \big( (k_n +\theta_n) \eps_\cN \big) + \int_{\p \cN} \vartheta^{EH}
    \label{nullsymp1}
\ee
where

\be
    \vartheta^{EH} = -\frac{1}{16 \pi} (i_{\d n^\m} \eps_\cN + \eps_{\cN 
 \n \rho}^\m \d n_\m) 
    \label{varthetaEH}
\ee
The Noether charge associated to the Einstein-Hilbert sympelctic potential \eqref{nullsymp1} is the well-known Komar charge 

\begin{equation}
    (q^{EH}_\xi)_{\m \n} = -\frac{1}{16 \pi} {\eps_\mathcal{M}}_{\m \n \rho \sigma} \nabla^\rho \xi^\sigma
    \label{komar}
\end{equation}
The master equation \eqref{master equation} allows us to relate the variation of the Komar charge for diffeomorphisms $\xi$ tangent to $\cN$ to the Einstein-Hilbert flux \eqref{nullsymp1} contracted with the field space vector $X_\xi$. In other words, we have 
\begin{equation}
    T_\n^\m \xi^\n {\eps_\mathcal{M}}_{\m} + dq_\xi^{EH} = I_\xi \theta^{EH} - i_\xi L^{EH}
\end{equation}
However, there are many terms involved in \eqref{nullsymp1}, and not all of them have an independent physical relevance. Ideally, we want to equip the null hypersurface $\cN$ with some boundary structure that allows us to identify the physical degrees of freedom. The boundary structure of a null hypersurface \cite{chandrasekaran2018symmetries} is given by the equivalence class of the triplet $(n^\m, n_\m, k_n)$, such that $(n^\m, n_\m, k)$ and $(n'^\m, n'_\m, k_{n'})$  belong to the same equivalence class if and only if there exists some function $A$ such that

\begin{align}
    n'_\m &= A n_\m \nn \\
    n'^\m &= A n^\m \nn \\
    k_{n'} &= Ak_n + n^\m \p_\m A
    \label{eqclass}
\end{align}
Hence, as the equivalence class $[n^\m, n_\m, k]$ describes the boundary structure of any null hypersurface of topology $B \times \mathbb{R}$, it is universal and should not give us any information on the physics on the boundary. Hence, for consistency, we have to restrict ourselves to the variations in phase space such that this boundary structure is preserved, i.e, to

\begin{align}
\d n^\m &= 0 \nn \\
\d n_\m &= 0 \nn \\
\d k_n &= 0
\label{conddefd0}
\end{align}
However, the quantities defined in \eqref{conddefd0} are one forms in field space, so they act on vectors in field space. Hence, we have to restrict ourselves to vectors $X_\xi$ in field space such that their contraction on the field space forms \eqref{conddefd0} vanishes. The diffeomorphisms $\xi$ tangent to the null hypersurface $\cN$ from which the field space vector $X_\xi$ is built are the ones preserving the universal structure of the null hypersurface, as explained in \cite{chandrasekaran2018symmetries} and in Appendix \ref{AppB}

\be
    \xi = (T(x^B) + v W(x^B)) \p_v - u W(x^B) \p_u + Y^A(x^B) \p_A
    \label{generaldiffeo0}
\ee
where the components $T(x^B)$, $W(x^B)$ and $Y^A(x^B)$ are respectively the affine supertranslations, the Weyl supertranslations, and the superrotations. Furthermore, we kept the term $- u W(x^B) \p_u$ in the expression of the vector field, even if it vanishes on the null hypersurface $\cN$, so it is not relevant for the restriction of $\xi$ on $\cN$. It is because the conditions \eqref{conddefd0} also fix this extension through the equations $l^\m \d_\xi n_\m = 0$ and $l_\m \d_\xi n^\m = 0$.

\vspace{0.3 cm}

The vectors fields \eqref{generaldiffeo0} are all tangent to the null hypersurface $\cN$ and so preserve its location. However, if the null hypersurface $\cN$ has a boundary $\p \cN$, then we must restrict the  group of infinitesimal diffeomorphisms preserving the boundary $\p \cN$, i.e schematically $\pounds_\xi \p \cN = 0$. For instance, if $\p \cN = S$ is some corner with topology $B$ that crosses each null generator one, then we should restrict ourselves to the vector $\xi$ that preserve $S$. As $S$ crosses every generator once, we can locate it at $v = 0$. So the group of diffeomorphisms preserving $\p \cN$ and the boundary structure \eqref{conddefd0} is then
\be
    \xi = W(x^B)(v\p_v - u\p_u) + Y^A(x^B) \p_A
    \label{grouppresbound}
\ee
which is \eqref{generaldiffeo0} without the affine supertranslations $T\p_v$, which move obviously the location of the corner.

\vspace{0.3 cm}

The vector fields spanned by \eqref{generaldiffeo0} form a closed algebra. The group obtained from the algebra of vectors \eqref{generaldiffeo} is 
\be
    Diff(S) \ltimes \mathbb{R}_W^S \ltimes \mathbb{R}_T^S
\ee
It is called the BMSW algebra at null infinity \cite{Freidel:2021fxf}, extending the famous BMS group of symmetry \cite{sachs1962asymptotic}. This procedure is not the original way of recovering this algebra. Instead, here we emphasized on the infinitesimal diffeomorphisms preserving the restrictions \eqref{conddefd0}, which gives us a more physical symplectic potential by killing some gauge redundancy. Indeed, if we impose \eqref{conddefd0}, we see that the Einstein-Hilbert bare potential \eqref{nullsymp1} becomes

\be
    \Theta^{EH} = \frac{1}{16 \pi} (\sigma_n^{\m \n} - \frac{1}{2} \theta_n \g^{\m \n}) \d \g_{\m \n} \eps_{\cN} + 2 \d (\theta_n \eps_\cN)
    \label{nullsymp2}
\ee
The expression of \eqref{nullsymp2} depends on the shear, the expansion and the volume form. We got rid of the unaffinity, the normal and the auxiliary vector. In addition, from \eqref{defexpansion}, we notice that the second term of \eqref{nullsymp2} is not only a boundary term in field space, but also in spacetime. Hence, we can obtain another symplectic potential $\Theta^D$ thanks to \eqref{generalsymppot}, by taking $l = -2 \theta_n \eps_\cN$ \footnote{the index "D" is a reference for Dirichlet, because the flux is in a Dirichlet form \cite{Odak2022}.}  

\be
    \Theta^D = \pbi{\Theta}^{EH} - d \d \frac{\eps_S}{8 \pi} = \frac{1}{16 \pi} (\sigma_n^{\m \n} - \frac{1}{2} \theta_n \g^{\m \n}) \d \g_{\m \n} \eps_\cN
    \label{Dirichletflux0}
\ee
The symplectic form obtained from \eqref{Dirichletflux0} is the same as the one obtained from \eqref{nullsymp2}, we have $\omega = \d \Theta^{EH} = \d \Theta^D$. We observe that the shear is conjugated to the conformal metric while the expansion is conjugated to the null volume form. Both shear and expansion characterize the intrinsic and extrinsic geometry of $\cN$, and the conformal metric is the data required by Sachs \cite{sachs1962characteristic} \footnote{The expansion is also needed at the intersection of both null hypersurfaces.} on the whole hypersurface $\cN$ for the initial value problem from a pair of two intersecting null hypersurfaces. Furthermore, this symplectic potential is covariant and vanishes for arbitrary variation $\d \g_{\m \n}$ around a shear free and expansion free null hypersurface, taken as a class of stationary solutions \cite{ashtekar2022charges}. Hence, as claimed in \cite{chandrasekaran2018symmetries}, it is the unique symplectic potential obtained from the Wald-Zoupas procedure \cite{wald2000general}\footnote{See \cite{Odak:2022ndm} for a modern review about the Wald-Zoupas procedure}. The charge associated to $\xi$ pullback on the cross section $S$ of constant parameter $v$ is given by the formula \eqref{generalcharged} and is equal to

\be
    Q^D_\xi(S) = \frac{1}{8 \pi}\int_S [W - \frac{1}{2} P_A Y^A - \theta_{\p_v}(T + v W)] \eps_S
    \label{Dirichlet charge}
\ee
which is the improved Noether charge \cite{Shi:2020csw} associated to the new symplectic potential \eqref{Dirichletflux0}. This is also the Wald-Zoupas charge. On a non-expanding horizon, $\sigma^{\m \n}_n = 0$ and $\theta_n = 0$ so $\Theta^D = 0$ for arbitrary variations, and the charges are conserved. 

\subsubsection{Local balance equation and entropy}

Now we come back to the master equation \eqref{master equation}. As $\g_{\m \n}$ is anomaly free, the symplectic potential \eqref{Dirichletflux0} contracted with a diffeomorphism of the symmetry group \eqref{generaldiffeo0} is of the form $P \pounds_\xi Q$, where $Q$ and $P$ are canonical pairs. Therefore it can be understood as a gravitational flux, vanishing if the momenta $P$ vanish or if the configuration space dynamical fields $Q$ remain unchanged when transported along $\xi$. Then we have, on a portion $\Delta \cN$ of $\cN$ between two cross sections $S_1$ and $S_2$ of $\cN$

\begin{align}
    \Delta Q^D_\xi &= \int_{\Delta \cN} I_\xi \Theta^D + \int_{\Delta \cN} T_{\m \n} \xi^\m n^\n \eps_{\Delta \cN} \nn \\
    &= \frac{1}{16 \pi} \int_{\Delta \cN} (\sigma_n^{\m \n} - \frac{1}{2} \theta_n \g^{\m \n}) \pounds_\xi \g_{\m \n} \eps_\cN + \int_\cN T_{\m \n} \xi^\m n^\n \eps_\cN
    \label{Dirichletflux}
\end{align}
Now, we consider only the subgroup of diffeomorphisms $\xi$ tangent to the null geodesics, i.e the supertranslations. We can see \eqref{Dirichlet charge} that the charge associated to the affine supertranslations vanishes on a non-expanding horizon. Furthermore, we know that the horizon null Killing field is a Weyl supertranslation and not an affine supertranslation, so we focus on Weyl supertranslations for now. If the null energy conditions are imposed, the creation term in the master equation becomes positive and the $Q^D_\xi$ variation is similar to a balance law for entropy, because the contribution of the matter degrees of freedom to the gravitational charge variation is always positive. In thermodynamics, entropy is defined at equilibrium and perturbatively near equilibrium. In general, when we have non infinitesimal gravitational flux and entropy creation terms, the charge $Q_\xi^D$ gives us a dynamical and local notion of entropy. We should recover the usual notion of entropy on the stationary solutions, which are the non expanding horizons. On the non-expanding horizon, thermodynamic equilibrium is achieved and the charges do not vary. However, the analysis of the perturbative non expanding horizon gives dynamical correction to the entropy at first order. 

\vspace{0.3 cm}

Let us assume that the unperturbed black hole is a stationary Kerr-Newman black hole with mass $M$, angular momentum $J$, electric charge $Q$ and area $A$. We can slightly perturb this stationary solution by introducing some (possibly charged) matter fields $\phi$, with corresponding stress energy tensor $T_{\m \n}$ such that

\begin{align}
    \phi = O(\eps) \nn \\
    T_{\m \n} = O(\eps^2)
    \label{smallpert}
\end{align}
where $\eps$ is a small quantity, where the meaning of "small" will be defined in the following. The background Killing vector is $\xi \stackrel{\cN}= \kappa v \p_v = \kappa v n$ on the dynamical horizon $\cN$, where $\kappa$ is chosen to be the black hole surface gravity \footnote{With this choice, it is well known that we can decompose the Killing field $\xi$ at infinity into a timelike Killing vector field normalized to $-1$ and a spacelike Killing vector field which generates closed orbits of length $2 \pi$.}. It is a Weyl supertranslation being part of the symmetry group \eqref{generaldiffeo0} preserving the boundary structure with parameter $W = \kappa$. Therefore, $\xi$ is null on the dynamical horizon and is exactly Killing when the black hole settles down to a stationary state, i.e in the far future. Thus, \eqref{smallpert} combined with the Einstein equations tells us that

\begin{align}
    \sigma_{n}^{\m \n} &= O(\eps^2) \nn \\
    \theta_n &=  O(\eps^2) \nn \\
    \pounds_\xi \g_{\m \n} &= O(\eps^2)
    \label{pertrubationstath}
\end{align}
implying

\be
    I_\xi \Theta^D = O(\eps^4)
    \label{thetaorderepsfour}
\ee
and so the parameter $\eps$ must be sufficiently small so that the metric perturbations of second order are negligible with respect to the first order metric perturbations of the Killing background. Now, from \eqref{Dirichlet charge}, \eqref{Dirichletflux} and \eqref{thetaorderepsfour}

\be
    \frac{\kappa}{8 \pi} \Delta(A - v\frac{dA}{dv}) = \int_{\cN} T_{\m \n} \xi^\m n^\n \eps_\cN + O(\eps^4) = S_c \geq 0
    \label{ppfldirichlet0}
\ee
which is equivalent to (see \cite{Rignon-Bret:2023lyn} for details)

\be
    \frac{\kappa}{8 \pi} \Delta(A - v\frac{dA}{dv}) = \Delta M - \Omega_H \Delta J - \Phi_{H} \Delta Q
    \label{ppfldirichlet}
\ee
and so the dynamical entropy is given by

\be
    S^D = \frac{1}{4}(A - v \frac{dA}{dv})
    \label{Dirichlet entropy 2}
\ee
This is a particular case of the dynamical entropy introduced in \cite{Wald2023} and \cite{Visser2023} for more general theories of gravity. It reduces to the usual Bekenstein Hawking entropy in the stationary case. It might seem surprising at first to not recover the usual physical process first law (PPFL)

\be
    \frac{\kappa}{8 \pi} \Delta A = \Delta M - \Omega_H \Delta J - \Phi_{H} \Delta Q 
    \label{classicalppfl}
\ee
but we should remember that we had to integrate between the bifurcation surface located at $v = 0$ and $v = + \infty$ to find \eqref{classicalppfl} \cite{wald1994quantum, gao2001physical, Rignon-Bret:2023lyn}. Here, we integrated between two arbitrary slices of constant $v$, so these terms remain. Hence, we can locally define an entropy variation in a physical process without talking about the bifurcation surface or the equilibrium state. In this process, the entropy creation term is of order $O(\eps^2)$ and the flux is of order $O(\eps^4)$, so the thermodynamic transformation is meant to be adiabatic. 

\vspace{0.3 cm}

Furthermore, if there is no matter crossing the dynamical event horizon on the portion $\Delta \cN$, the Raychaudhuri equation gives us that $\p_v \theta_{\p_v} = - \theta_{\p_v}^2 - \sigma_{\p_v, \m \n} \sigma_{\p_v}^{\m \n}$, so $\theta_{\p_v} = O(\eps^4)$ and in this case

\be
    T_H \Delta S^D = \frac{\kappa}{8 \pi}\int_{\Delta \cN} \sigma_{n,\m \n} \sigma_n^{\m \n} \eps_{\cN} + o(\eps^4) = \mathcal{Q}^D \geq 0
\ee
is a local entropy variation. Thus the entropy variation around a non expanding horizon is positive up to second order. If there is no matter on $\Delta \cN$, the only piece contributing to the entropy variation is the heat current which is a second order term. This term can be interpreted as the energy flux of the weak gravitational waves crossing a perturbed non expanding horizon \cite{ashtekar2022charges}. Hence, the energy carried by weak gravitational waves is the heat flux contributing to the entropy variation. 

\vspace{0.3 cm}

The presence of a Killing field which has a timelike Killing component normalized to $-1$ at infinity
establishes a well defined notion of temperature seen by a fay away observer, the Hawking temperature. However, if we are not close to equilibrium, there is no well-defined intrinsic notion of temperature on the null hypersurface $\cN$. Even for a black hole at equilibrium, the Hawking temperature makes sense for a far away observer, but an observer accelerating near the black hole horizon observes a different temperature. Indeed, in order to derive the Physical Process First Law (PPFL) from \eqref{Dirichletflux}, we chose to take $\xi$ as a Weyl supertranslation with parameter $W(x^A) = \kappa$ on the null horizon, where $\kappa$ is the surface gravity of the background black hole. Nevertheless, when the portion $\Delta \cN$ of the null hypersurface $\cN$ we are interested in is not a (perturbed) stationary event horizon, there is no canonical choice of $\kappa$. As it is well known, $\kappa$ is chosen as the surface gravity of the stationary black hole because $\xi$ is meant to be identified with is the Killing field generating the null horizon with timelike part normalized to $-1$ at infinity. However, any Weyl supertranslation $\xi = W (v \p_v - u \p_u)$ still preserves the boundary structure \eqref{generaldiffeo0}. Even if the temperature on the null hypersurface is not defined because of the absence of a global Killing vector in general, we can however identify in sufficiently small regions observers of local constant acceleration $W$ moving along the lines of tangent vector $\xi$, close to $\cN$. This result is true for large enough local acceleration $W$, see Appendix.\ref{AppC}. These observers measure a local Unruh temperature, and so there exists an analogy between Weyl supertranslation diffeomorphisms and monothermal thermodynamic transformations \footnote{Reminder : Monothermal does not mean isothermal. It means that the environment is at constant temperature during the thermodynamic transformation, but the system is not. Indeed, when the system is not in internal equilibrium, its temperature is generally not well defined. Here we associate the Weyl supertranslation transformation to monothermal thermodynamic transformations because the local observer sees locally a surrounding thermal bath of temperature $\frac{\kappa}{2\pi}$ but the system itself has not a well defined temperature in general.}. In this case, the temperature is a property of the field $\xi$ and the trajectory of (approximate) constant acceleration we choose, and not a property the gravitational system. Indeed, the local balance law \eqref{Dirichletflux} is independent of the parameter $W$, and we always can normalize $\xi$ such that $W = 1$.

\subsection{Charge variation on a perturbed Killing horizon}

Hawking, Perry and Strominger argued that information can be stored in a holographic manner on the black hole event horizon due to shifts of the null generators caused by ingoing particles crossing the surface \cite{Hawking:2015qqa, Hawking:2016msc, Hawking:2016sgy}. The supertranslations preserve the background structure of any null hypersurface with topology $B \times \mathbb{R}$ because the null geodesics are independent of each other, they never cross \footnote{On a general black hole event horizon, the generators can never leave the horizon and never cross too. However, they can enter the horizon at points called caustics.}. Hence, any local disturbance of a bunch of geodesics cannot affect the other geodesics. If we use the general flux balance law \eqref{Dirichletflux} between two cross sections on a perturbed black hole horizon, we get at first order
\be
    \D Q_\xi^D = \int_{\Delta \cN} T_{\m \n} \xi^\m n^\n \eps_\cN + O(\eps^4)
\ee
for any $\xi$ belonging to the BMSW algebra. As $\xi$ is a general Weyl supertranslation, we can use the arbitrariness of the parameter $W(x^A)$ to write balance laws at fixed angular direction $x^A$. Hence, the variation of the charges $Q_{W v \p_v}$ depend directly on the details of the stress energy falling into the black hole, and so the charges store some information about the collapsed matter which formed the black hole. On the stationary event horizon, on any cross section $S$, the charge is given by 
\be
    Q_{Wv\p_v} = \int_S W(x^A) \eps_S
\ee
and so the local area element $\eps_S$ is the observable where the information about the local density of stress energy which fell into the black hole is stored. A similar analysis can be held for the superrotations.

\subsection{Local entropy balance law for York flux}

\subsubsection{Legendre transformation}

Before getting to the heart of the matter, it is worth spending some time on well known notions in order to understand better what we are doing in the following. In thermodynamics, the second law, or the entropy balance law, can generally be written as

\begin{equation}
\begin{aligned}
    dS &= \frac{\mathcal{Q}}{T_{ext}} + S_c \\
    &= \frac{dE + P_{ext} dV - \cdot \cdot \cdot}{T_{ext}} + S_c
    \label{ptS}
\end{aligned}
\end{equation}
with $S_c \geq 0$ and where we used the first law $dE = \mathcal{W} + \mathcal{Q}$ to go from the first to the second line. In general we could have added any kind of external work $\mathcal{W}$. For instance, if we are for instance in a system with total electric charge $Q$ and external electrostatic potential $\Phi_{ext}$, or with a number $N_i$ of particles of type $i$ with external chemical potential $\mu_{ext, i}$, we can write

\begin{equation}
    \mathcal{W} = - P_{ext} dV + \Phi_{ext} dQ + \sum_i \mu_{ext, i} dN_i
    \label{work}
\end{equation}
However, keeping only $\mathcal{W} = - P_{ext} dV$ in the above formulas is enough to illustrate our purpose. Under these circumstances, the flux term vanishes if we set $dE = dV = 0$, and then we get $dS = S_c \geq 0$. Therefore, for systems with constant energy and constant volume (microcanonical ensemble), $S$ is identified as the thermodynamic potential, because its variation is always positive and vanishes only at equilibrium, so it gives an indication about the spontaneous evolution of the system.

\vspace{0.3 cm}

However, if we proceed to a Legendre transformation and set now $F = E - T_{ext} S$ \footnote{Here, $F$ is not exactly the free energy, as in general $T_{ext} \neq T$. In fact it is not a state function as $T_{ext}$ is not the temperature of the system which may not be well defined.}, we obtain from \eqref{ptS}

\begin{equation}
    - \frac{dF}{T_{ext}} =  \frac{S dT_{ext} + P_{ext} dV}{T_{ext}} + S_c
    \label{freeenergy}
\end{equation}
Here the flux term vanishes when $dT_{ext} = dV = 0$, and therefore $-dF \geq 0$ when the external temperature and the volume of the system are fixed during the physical process (canonical ensemble). In that case, $F$ is the appropriate thermodynamic potential. The point here is that the good thermodynamic potential depends on the physically motivated form of the flux. Similarly to \eqref{ptS} and \eqref{freeenergy}, the master equation \eqref{master equation} relates the charge variation to some flux and a positive term if the null energy conditions are imposed. From a given bulk Lagrangian, $\Theta$ is defined up to exact terms in spacetime and field space \cite{jacobson1994black}. 

\vspace{0.3 cm}

In the previous section, we used a symplectic potential written in a Dirichlet form \ref{Dirichletflux}, as in \cite{chandrasekaran2018symmetries}, and worked with the corresponding improved Noether charges. However, even if \eqref{Dirichlet entropy 2} allows us to recover the usual PPFL locally, it cannot give a satisfactory global notion of entropy far from equilibrium. Indeed, for a Schwarzschild black hole formed after a spherical collapse for instance, the event horizon is initially a light cone in Minkowski spacetime bent by the gravitational effects of the collapsing matter (see Fig.\ref{Figurelightcone}). However, initially, when spacetime is still flat, the entropy \eqref{Dirichlet entropy 2} is negative and decreases as we can check by using \eqref{Dirichletflux}. Even if the entropy can decrease for open physical systems, it seems unnatural for it to vary on the Minkowski's light cone. Indeed, it is embedded in flat spacetime and we do not expect that the gravitational charges evaluated on its cross sections vary because the cancellation of the Weyl tensor means that the gravity degrees of freedom are not excited. If we understand entropy as a gravitational charge, we may expect that it vanishes on any cross section of the light cone embedded in flat spacetime. We will define such an entropy, with vanishing flux on the Minkowsk's light cone and on non-expanding horizon. In other words, we want these two portions of $\cN$ to be \textit{stationary}, in the sense that all the gravitational charges associated to the diffeomorphisms preserving the boundary structure of a general null hypersurface do not vary. Furthermore, this new entropy vanishes on Minkowski  light cone and gives the usual Bekenstein-Hawking entropy on a non-expanding horizon \footnote{Minkowski spacetime and stationary black hole spacetimes both possess a Killing field that is timelike Killing field when it approaches $\cN$. They are stationary in that sense.}. It increases on a spherically symmetric cross sections of any spherically symmetric outgoing null hypersurface, and so in particular for the event horizon formed through a spherically symmetric collapse.

\subsubsection{York flux and charges and algebra}

In order to do so, we start from the analysis presented in \cite{Odak2022}. In this paper, alternative boundary condition on the null hypersurfaces are presented. One possible symplectic potential was the Dirichlet like symplectic potential \eqref{Dirichletflux} that can also be written as

\be
    \Theta^D = \frac{1}{16 \pi} ( \sigma_n^{\m \n} \d \g_{\m \n} \eps_\cN - \theta_n \d \eps_\cN)
    \label{Dirichlet flux2}
\ee
using the useful identity

\be
    \d \eps_\cN = \frac{1}{2} \g^{\m \n} \d \g_{\m \n} + l_\m \d n^\m
    \label{identityepsgamma}
\ee
and imposing $\d n^\m = 0$ in order to preserve the boundary structure. Then we can integrate by part \eqref{Dirichlet flux2} $- \theta_n \d \eps_\cN $ to $\eps_\cN \d \theta_n$. Hence we have now the symplectic potential

\be
    \Theta^Y = \frac{1}{16 \pi}( \sigma_n^{\m \n} \d \g_{\m \n} \eps_\cN + \eps_\cN \d \theta_n) = \pbi{\Theta}^{EH} - d \d \frac{\eps_S}{16 \pi} = \Theta^D + d \d \frac{\eps_S}{16 \pi}
    \label{Yorksumppot}
\ee
that we can name the York symplectic potential, as the phase space variables are the conformal metric $\hat{\g}_{\m \n}$ and the expansion $\theta_n$ \footnote{The shear is tracefree, so $\sigma^{\m \n} \d \g_{\m \n} = \sigma^{\m \n} \d \hat{\g}_{\m \n}$}. These are the Sach's free data \cite{sachs1962characteristic}\footnote{For the initial value problem of two null hypersurfaces intersecting at some corner, we need to know the conformal metric on the null hypersurfaces and the expansions of both null hypersurfaces at the corner, in addition to the bracket between the normal $n$ and the auxiliary vector $l$ at the corner. As here we are interested in only one of the two null hypersurfaces, the relevant data are $\hat{\gamma}_{\m \n}$ on $\cN$ and $\theta_n$ at the corner. The Raychaudhuri equation gives $\theta_n$ everywhere on $\cN$ from the value of $\theta_n$ at the corner and the shear $\sigma_{\m \n}$ that can itself be obtained by taking the tracefree Lie derivative of the conformal metric.}. The main point of Sach's analysis is precisely that, on null hypersurfaces, we know exactly what are the physical degrees of freedom, and what is gauge. Hence expressing symplectic potential with canonical variables $(\hat{\g}_{\m \n}, \theta_n)$ is quite natural. Furthermore, \eqref{Yorksumppot} contracted with one of the general diffeomorphisms \eqref{generaldiffeo0} preserving the boundary structure gives

\be
    I_\xi \Theta^Y = \frac{1}{16 \pi}( \sigma_n^{\m \n} \pounds_\xi \g_{\m \n} \eps_\cN + \eps_\cN \pounds_\xi \theta_n + \eps_\cN \Delta_\xi \theta_n)
    \label{Yorkfluxanom}
\ee
and the associated improved Noether charge (integrated on $S$) can be deduced from \eqref{generalcharged} is 

\be
    Q^Y_\xi(S) = \frac{1}{8 \pi}\int_S [W - \frac{1}{2} P_A Y^A - \frac{1}{2}\theta_{\p_v}(T + v W)] \eps_S
    \label{York charge}
\ee
If we look at \eqref{Yorkfluxanom}, we notice that it is not in the form $P \pounds_\xi Q$ unlike \eqref{Dirichlet flux2}. Indeed, while $\g_{\m \n}$ is anomaly free, it is not the case for $\theta_n$, and the term $\Delta_\xi \theta_n$ does not vanish in general. This is because $\theta_n$ is not class III invariant \footnote{Remember that a class III invariant quantity is a physical quantity that is invariant through a rescaling of the normal $l$ and the auxiliary vector $n$ preserving the relation $n_\m l^\m = -1$, i.e invariant through $(n, l) \longrightarrow (An, A^{-1} l)$.} in the sense of Chandrasekhar \cite{chandrasekhar1998mathematical}, as $\theta_{An} = A \theta_n$, see Appendix \ref{AppD} for more details. The anomaly depends on the chosen representative $n$. Hence, to get rid of this anomaly term in the computations \footnote{The formula for the flux is class III invariant though. However, in general we cannot write it with Lie derivatives only.} we choose a preferred normal, giving non anomalous contributions to the flux

\be
    n^\m= v (\frac{\p}{\p v})^\m
    \label{normalyork}
\ee
and which gives us $k_n = 1$. It is shown in Appendix.\ref{AppD} that the subgroup of diffeomorphisms \eqref{generaldiffeo0} which are non anomalous is given by the following vectors

\be
    \xi = W(x^B) (v \p_v - u \p_u) + Y^A(x^B) \p_A
    \label{generaldiffeo00}
\ee
 which have a closed algebra. The Lie group associated to this algebra is 
 \be
    Diff(S) \ltimes \mathbb{R}_W^S
 \ee
 The subalgebra spanned by the vectors \eqref{generaldiffeo00} is exactly the subalgebra \eqref{grouppresbound} that preserves the location of some corner $S$ of $\cN$ that cross any null generator once, and so it is the symmetry group of diffeomorphsims preserving the location of $\cN$, of its boundary $\cN$ and the boundary structure \cite{chandrasekaran2018symmetries,Chandrasekaran:2019ewn}. In particular, they preserve the boundary of the ingoing and outgoing light cones. Therefore, with this restricted choice of diffeomorphisms, the flux \ref{Yorkfluxanom} and the improved Noether charge \ref{York charge} become respectively

 \be
    I_\xi \Theta^Y = \frac{1}{16 \pi} (\sigma_n^{\m \n} \pounds_\xi \g_{\m \n} \eps_\cN + \eps_\cN \pounds_\xi \theta_n)
    \label{Yorkflux}
 \ee
 and

\be
    Q^Y_\xi(S) = \frac{1}{8 \pi}\int_S [W(1 -  \frac{\theta_{v \p_v}}{2}) - \frac{1}{2} P_A Y^A] \eps_S
    \label{Yorkusecharge}
 \ee
 Hence we get a new flux written in the form $P \pounds_\xi Q$, and new charges \cite{Odak2022}. This flux and this charge are similar to the one introduced in \cite{hopfmuller2018null}, but we restricted ourselves to the covariant phase space introduced in \cite{chandrasekaran2018symmetries} and get charges linear in the parameters $W(x^B)$ and $Y^A(x^B)$. The physical motivations to introduce them are also quite different. 
 However, it is worth emphasizing that the symplectic potential \eqref{Yorksumppot} does not satisfy the Wald-Zoupas requirements. If it is indeed covariant with respect to the diffeomorphisms \eqref{generaldiffeo00}, it does not satisfy the Wald-Zoupas stationary solution requirement, i.e there is no so called stationary solution $\phi = (\hat{\g}_{\m \n}, \theta_n)$ such that $\Theta^Y(\phi, \d \phi)$ vanishes for arbitrary variations $\d \phi$ \cite{wald2000general, Odak:2022ndm}. However, in order to build a vanishing Noether flux on Minkowski light cone, we have to go beyond the Wald-Zoupas procedure and accept as a suitable flux $I_\xi \Theta^Y$ which vanishes for any allowed symmetry of the boundary structure $\xi$, the diffeomorphisms \eqref{generaldiffeo00}. Within this definition of a stationary solution, the symplectic flux and the associated charges both vanish on Minkowski's light cone. Indeed, on  Minkowski's light cone $\sigma_n^{\m \n} = 0$. Furthermore, the outgoing null light cone is defined as a null hypersurface $u = 0$, and the affine parameter $v$ goes from $v = 0$ (the light cone tip) at $r = 0$ to $v = + \infty$. Thus, for the Minkowski light cone, we have $r = v$ for any value of the affine parameter $v$. Hence, on sections of constant $v$ (or constant $r$)

 \be
     \theta_n = v \theta_{\p_v} = \frac{v}{\d A} \frac{d \d A}{dv} = \frac{r}{\d A} \frac{d \d A}{dr} = D - 2
     \label{vexpansion}
 \ee
 and so $\theta_n = 2$ in $D = 4$ dimensions. Therefore, on the outgoing Minkowiski light cone for which $P_A = 0$ \footnote{In fact, the angular momentum aspect is equal to the normalized twist $\eta_A = - \g_\m ^\n \p_u^\rho \nabla_\n \p_\rho u$ on the null hypersurface, in fact $\eta_A = - \frac{1}{2} P_A$. The word normalized is added here because the twist is not a class III invariant quantity as we can easily check, see \cite{Odak2022} for more details.} and $\theta_n = 2$, we get from \eqref{Yorkflux} that $I_\xi \Theta^Y = 0$ and from \eqref{Yorkusecharge} $Q_\xi^Y(S) = 0$ for any $\xi$ that belongs to the symmetry group of the boundary structure \eqref{generaldiffeo00}. Hence, we have a vanishing flux and vanishing charges on Minkowski light cone, as desired \footnote{There are other notions of entropy that have been introduced on the Minkowski light cone in order to simulate analogies with black hole thermodynamics. In particular, in \cite{de2018light, de2019light},  the entropy is given by the \textit{conformal} area at first order, and a similar procedure to the one occurring here is done in order to remove the order $0$ expansion of the Minkowski's future light cone. However, the entropy proposed in these paper is obtained using the assumption that the first order expansion vanishes at infinite affine parameter $v$ on the hypersurface, so it does not equal our \eqref{Yorkentropy}. Furthermore, the associated temperature is not associated to a boost Killing field but to the radial special conformal field.}. We also notice that the flux \eqref{Yorkflux} vanishes on non expanding horizons, because $\sigma_n = 0$ and $\theta_n = 0$ everywhere. 
 
\vspace{0.3 cm}

 In addition, we can introduce the symplectic form $\omega = \d \Theta^{EH} = \d \Theta^{D} = \d \Theta^{Y}$ and get
 \be
    - I_\xi \omega = d(\d q_\xi^Y - i_\xi \Theta^Y)
        \label{sympformform}
 \ee
 on-shell. So the charge $Q_\xi^Y$ is Hamiltonian if $\Theta^Y = 0$ or if $\xi$ is tangent to the corner (the former can be achieved if the boundary conditions $\d \hat{\g}_{\m \n} = \d \theta_n = 0$ are satisfied). Furthermore, it is very important to keep in mind that the York symplectic potential $\Theta^Y$ is covariant \footnote{The Dirichlet symplectic potential $\Theta^D$ is covariant too.}, which means it satisfies 
 \be
    (\d_\xi - \pounds_\xi) \Theta^Y = 0
    \label{covariance}
 \ee
This is because it differs from the Einstein-Hilbert symplectic potential by a boundary term proportional to the exterior derivative of the area, and which is covariant as well. Let $\xi_1$ and $\xi_2$ be two field independent diffeomorphisms belonging to the algebra \eqref{generaldiffeo00}. Using the relations $[\pounds_{\xi_1}, I_{\xi_2}] = 0$ and $[\d_{\xi_1}, I_{\xi_2}] = - I_{[\xi_1, \xi_2]}$, we find that on-shell (in vacuum) \footnote{On-shell, the Einstein-Hilbert Lagrangian vanishes and so we do not need to include this term in the following equations.}
\begin{align}
    d \Delta_{\xi_2} q_{\xi_1}^Y =(\d_{\xi_2} - \pounds_{\xi_2}) I_{\xi_1} \Theta^{Y} &= - I_{[\xi_1, \xi_2]} \Theta^Y  \nn \\
    &= - d q_{[\xi_1, \xi_2]}^Y 
\end{align}
Therefore if we contract \eqref{sympformform} with another diffeomorphism we get 
\begin{align}
    - I_{\xi_1} I_{\xi_2} \omega &= d(\d_{\xi_1} q_{\xi_2}^Y - i_{\xi_2} I_{\xi_1} \Theta^Y) \nn \\
    &= d ( q_{[\xi_1, \xi_2]}^Y + \pounds_{\xi_1} q_{\xi_2}^Y - \pounds_{\xi_2} q_{\xi_1}^Y)
\end{align}
Now we can consider a spacelike hypersurface $\Sigma$ crossing the boundary $\cN$ at the corner $S$. We can therefore define the following Poisson bracket 
\be
    \{ Q_{\xi_1}^Y, Q_{\xi_2}^Y \}^{BT} = - \int_\Sigma I_{\xi_1} I_{\xi_2} \omega - \int_S \pounds_{\xi_1} q_{\xi_2}^Y - \pounds_{\xi_2} q_{\xi_1}^Y = Q^Y_{[\xi_1, \xi_2]}
\ee
which is the Barnich-Troessaert Poisson bracket \cite{Barnich:2009se, Barnich:2010eb, Barnich:2011mi}, first introduced to compute the BMS algebra. The only required properties to obtain this Poisson bracket are the covariance \eqref{covariance} and being on-shell. Furthermore, on a non expanding horizon or on Minkowski light cone, $I_\chi \Theta^Y = 0$ and so we have that 
\be
    \{ Q_\chi^Y, Q_\xi^Y \}^{BT} = \int_\Sigma -I_\chi I_\xi \omega = \int_{S} \d_\chi Q_\xi^Y = - \int_{S} \d_\xi Q_\chi^Y
\ee
and so the charge $Q_\chi$ generates the symmetry transformation $X_\chi$ in phase space on the stationary solutions.

\subsubsection{General properties of the flux and positive flux theorems}
\label{generalpropertyflux}
 
With the new flux \eqref{Yorkflux} and the new charges \eqref{Yorkusecharge}, the master equation \eqref{master equation} becomes

 \begin{align}
    \Delta Q_\xi^Y &= \int_{\Delta \cN} I_\xi \Theta^Y + \int_{\Delta \cN} T_{\m \n} \xi^\m n^\n \eps_{\cN} \nn \\
    &= \frac{1}{16 \pi}\int_{\Delta \cN} \sigma_n^{\m \n} \pounds_\xi \g_{\m \n} \eps_\cN + \eps_\cN \pounds_\xi \theta_n + \int_{\Delta \cN} T_{\m \n} \xi^\m n^\n \eps_\cN
    \label{mastereqyork}
 \end{align}
If $\xi$ is a Weyl supertranslation, $\xi^\m = W(x^A) n^\m$ and the Raychaudhuri equation $\pounds_\xi \theta_n = W(x^A) v \p_v \theta_n = W(x^A) (\theta_n - \frac{1}{2} \theta_n^2 - \sigma_n^2 - R_{\m \n} n^\m n^\n)$ combined with the Einstein equations transforms \eqref{mastereqyork} into

\be
    \Delta Q_{Wn}^Y = \frac{1}{8 \pi}\int_{\Delta \cN} W\eps_\cN \bigg( \frac{\sigma_n^2}{2} + \frac{\theta_n}{\theta_{n0}} - \big( \frac{\theta_n}{\theta_{n0}}\big)^2 \bigg) + \frac{1}{2} \int_{\Delta \cN} W T_{\m \n} n^\m n^\n \eps_\cN
    \label{positiveflux}
\ee
where $\theta_{n0} = 2$ is the value of expansion on the outgoing Minkowski's light cone, considered as the reference (stationary) solution. The RHS of \eqref{positiveflux} is positive as long as the null energy conditions are satisfied, $W(x^A) > 0$ and $\theta_{n0} \geq \theta_n \geq 0$. This last condition is a very non trivial one, in the sense that it is not a priori physically relevant. Indeed, the value on $\theta_n$ depends on the extrinsic geometric properties of the considered null hypersurface. For a generic null hypersurface embedded in flat spacetime, it can take any value. Furthermore, the condition $\theta_n \leq 0$ does not seem to be physically relevant either \footnote{except when the cross section is a marginally trapped surface, in that case we also need no have $\theta_l \leq 0$. We know however that in general such a condition implies, through the Raychaudhuri equation, that the expansion diverges for a finite affine parameter, and so we cannot extend the affine parameter ton infinity. Hence the chosen geodesic congruence is not future complete.}  However, there are still some cases where the latter condition is relevant. 

\vspace{0.3 cm}

First, we can restrict ourselves to positive expansion null hypersurfaces, satisfying $\theta_n \geq 0$ everywhere. By doing so, we can avoid caustics, which necessarily form at some parameter $v > v_0$ if $\theta_n$ is negative at $v_0$. Second, we should notice that if $\theta_n(v_0) \leq 2$, then for any $v > v_0 > 0$, $\theta_n(v) \leq 2$ if the null energy conditions are satisfied (for fixed angular coordinates $x^A$). Indeed, if it is not true, there exists a parameter $v_P > v_0$ such that $\theta_n(v_P) > 2$. But the Raychaudhuri equation is 

\be
    v\p_v \theta_n = \theta_n - \frac{\theta_n^2}{2} - \sigma_n^2 - T_{\m \n} n^\m n^\n
    \label{Raychau1}
\ee
and so if $\theta_n \geq 2$ and if the null energy conditions are satisfied, $\p_v \theta_n$ is negative (as $v > v_0 > 0$) . It means that $\theta_n(v \leq v_P) > 2$, and so $\theta_n(v_0) > \theta_n(v_P) > 2$, and we get a contradiction. Of course, if $\theta_n(v_0) < 0$, then $\theta_n(v > v_0) < 0$ for the same reasons.  Hence, if for any $x^A$, $\theta_n(x^A, v = + \infty)$ exists and if there exists $v_0$ such that $\theta(x^A, v_0) \leq 2$, then $0 \leq \theta_n(x^A, [v_0, +\infty[) \leq 2$. Therefore, the charge $Q_{n}^Y$ is positive and increases on $[v_0, +\infty[$. Of course, as we already mentioned, we had to restrict ourselves to (portion) of null hypersurfaces $\cN$ which have topology $B \times \mathbb{R}$. In particular, these hypersurfaces should not allow some generators to enter or leave $\cN$, as it is the case on general event horizons where generators can enter $\cN$ at caustics. However, the boundary of the null hypersurfaces that we consider may contain caustics, as the outgoing light cone's tip. 

\vspace{0.3 cm}

As a consequence, an arbitrary null hypersurface $\cN$ with topology $B \times \mathbb{R}$ which is future complete has the following property. Take any spacelike compact cross section $S$ of $\cN$ which intersects all the null generators exactly once, look at the vector fields belonging to the symmetry group of $\cN$ which preserves the location of the corner $S$. Among them, there are the Weyl supertranslation vector fields which vanishes on $S$, i.e 
\be
    \left. \xi_W \right|_S = 0
\ee
For the following, we normalize $W$ to $W = 1$ and so the Weyl supertranslationt vector field $\xi_W$ is the normal $n = v\p_v$ on $\cN$ \footnote{However, the following theorem works for any $\xi = W(x^A) n$ with $W > 0$. In other words, $\xi$ must be future pointing in the future of $S$.}. As $\xi = n = v\p_v$, the corner $S$ is located at the value $v = 0$ of the affine parameter $v$. Then, for any compact spacelike cross sections $S_1$ and $S_2$ of $\cN$ in the future of $S$, such that $S_2$ is in the future of $S_1$ (which can be noted $S_2 \succeq S_1 \succeq S$), we have that 
\be
    \Delta Q_n^Y = Q_{n,S_2}^Y - Q_{n, S_1}^Y \geq 0
    \label{increasingcharge}
\ee
Indeed, from \eqref{positiveflux}, $Q_n^Y$ increases if the null energy conditions are satisfied as long as $\theta_{n0} = 2 \geq \theta_n \geq 0$. But as $\cN$ is future complete, we necessarily have that $\theta_n \geq 0$, because if it was not true, then from the Raychaudhuri equation, $\theta_n$ would become infinite at a finite value of an affine parameter $v$, and there would be a formation of a caustic. If null generators cross, the topology of $\cN$ is not $B \times \mathbb{R}$ so this is a contradiction. Hence $\theta_n \geq 0$. Furthermore, we have that $\left. \theta_n \right|_S = 0 $. Indeed, we chose to define the coordinate system such that the corner $S$ is located at $v = 0$. Hence, $\theta_{\p_v}$ does not diverge at $v = 0$, because no caustics form on $S$ \footnote{Of course, this is not the case for the light cone.}, and so $\theta_{v \p_v} (0, x^A) = \theta_n (0, x^A) = 0$. It implies that $0 \leq \theta_n(v > 0, x^A) \leq 2$ by the theorem discussed before. Thus $\theta_n \leq 2$ at any point on $\cN$ in the future of $S$. This proves \eqref{increasingcharge}. The charge associated to the normal $n$ on a cross section $S'$ of constant affine parameter $v > 0$ is then given by \eqref{Yorkusecharge} 
\be
    \left.  Q_n^Y \right|_{S'} = \frac{1}{8\pi} \int_{S'} (1 - \frac{1}{2} \theta_n) \eps_S = \frac{1}{8\pi}(A - \frac{v}{2} \frac{dA}{dv})
\ee
In addition, as $\left. \theta_n \right|_S = 0$, we have 
\be
    \left. Q_{n}^{Y} \right|_S =  \frac{1}{8\pi} \left. A \right|_S
\ee
and so on any cross section $S'$ in the future of $S$, $Q^Y_{n, S'} \geq \frac{1}{8\pi} \left. A \right|_S \geq 0$ and the dynamical entropy associated with the vector field $\xi = n$ is positive. This result can be generalized to any dimensions.

\subsubsection{Perturbation on Killing horizons}

On a slightly perturbed stationary black hole event horizon, $\theta_n$
 is arbitrary small and positive, because $\theta_n (v = + \infty) = 0$. Thus, of course, the right hand side of \eqref{positiveflux} is positive and we take $\xi \stackrel{\cN}= \kappa v \p_v$ as the background Killing field, with $\kappa$ being the surface gravity of the stationary black hole. At first order in perturbation \footnote{Remember that according to our conventions, a linear perturbation of the metric is of order $\eps^2$.}, we get from \eqref{mastereqyork} \footnote{We have to work out the $\eps_\cN \theta_n$ term
 \begin{align}
    \frac{1}{16 \pi}\int_{\Delta \cN} \eps_\cN \pounds_\xi \theta_n
    &= \frac{1}{16 \pi}\int_{\Delta \cN} \frac{dv d^2 x^A}{v} \sqrt{\g} \kappa v \p_v \theta_n \nn \\
    &=  \frac{1}{16 \pi}\int_{\Delta \cN} dv d^2 x^A \kappa \p_v (\sqrt{\g} \theta_n) - \frac{1}{16 \pi}\int_{\Delta \cN} \kappa \eps_\cN \theta_n^2 \nn \\
    &= \Delta (\frac{\kappa A \bar{\theta}_n}{16 \pi}) + O(\eps^4)
 \end{align}
 where the variation $\Delta$ is evaluated between two cross sections of constant $v$. } 

 \be
    T_H \Delta S^Y = \Delta (\frac{\kappa A}{16 \pi} \bar{\theta}_n) + \Delta M - \Omega_H \Delta J - \phi_H \Delta Q + O(\eps^4)
    \label{Firstlaw york}
 \ee
 where $\bar{\theta}_n = \frac{v}{A} \frac{dA}{dv}$ \footnote{In general we have $\theta_n = \frac{v}{\sqrt{\gamma}}\frac{d\sqrt{\g}}{dv} \neq \frac{v}{A}\frac{dA}{dv} = \bar{\theta}_n$. $\theta_n$ is local on the cross section while $\bar{\theta}_n$ is not. However, the equality holds in the spherically symmetric case.} and on cross sections of constant affine parameter $v$ the York entropy is given by

 \be
    S^Y = \frac{A}{4}(1 - \frac{\bar{\theta}_n}{2}) = \frac{1}{4}(A - \frac{v}{2} \frac{dA}{dv})
    \label{Yorkentropy}
 \ee
 It is worth emphasizing again that this entropy clearly vanishes on Minkowski's light cone and equals the Bekenstein-Hawking entropy on a non-expanding horizon. How can we interpret the additional term $\Delta (\frac{\kappa A }{8 \pi}\frac{\bar{\theta}_n}{\bar{\theta}_{n0}})$ in the PPFL ? It is important to remember that this is a physical process first law which is different from the equilibrium state version, as explained in the introduction. In thermodynamics, the law $dE = -PdV + T dS$ relates two nearby stationary solutions in phase space. This is strictly speaking an identity, relating state variables defined at equilibrium.  Here, we study a physical process, and we are not at equilibrium on any slice of constant $v$. This is why the entropy needs not to be the Bekenstein-Hawking entropy, but includes dynamical corrections. In that sense, it makes much more sense to write the balance law $dS = \frac{\mathcal{Q}}{T_{ext}} + S_c$ than the identity $dS = \frac{dM}{T} - \Omega_H \frac{dJ}{T} - \phi_H \frac{dQ}{T}$ on an arbitrary portion of $\cN$. Furthermore, while the Dirichlet flux vanishes at first order, it is not the case of the York flux, giving a non-vanishing gravitational flux. We expect in general a gravitational flux, even near equilibrium, because the geometry varies along the dynamical event horizon $\cN$. The Legendre transformation of the Dirichlet symplectic potential \eqref{Yorksumppot} enables us to construct some flux (the York flux) which takes into account the change of geometry on a portion of the null horizon where no matter crosses it.  

 \vspace{0.3 cm}
 
 However it does not mean that we cannot make sense of the entropy balance law for York potential for a perturbed non expanding horizon. First we have to notice that in vacuum, just before or after matter felt into the black hole, $T_{\m \n} \xi^\m n^\n \eps_\cN = T_H S_c = 0$, and we get from \eqref{mastereqyork} the linearized entropy balance equation \eqref{second law thermo}

 \be
    T_H \Delta S^Y = \mathcal{Q}^Y = \frac{\kappa}{16 \pi} \Delta (v \frac{dA}{dv}) + O(\eps^4) = \frac{T_H}{8} \Delta A + O(\eps^4)
    \label{PPFLvaccuumdim4}
 \ee
 where we used the linearization at first order in vacuum of \eqref{positiveflux}, and where $\mathcal{Q}^Y$ is the "heat flux" appearing in \eqref{second law thermo}, equal to the pullback of the Noether York current $j_\xi^Y = I_\xi \Theta^Y - i_\xi L^Y$ on the null horizon. It is worth noticing that since the perturbation $T_{\m \n}$ is of order $\eps^2$ \footnote{For the charged case the total stress energy tensor $T_{\m \n}$ is not of order $\eps^2$, so we have to add the Yang-Mills Lagrangian to the Einstein-Hilbert one and compute the total Noether current, which exterior derivative also vanishes at order $\eps^4$, as it is done in \cite{Rignon-Bret:2023lyn}.}, and since $\xi$ is a background Killing vector, then $j_\xi^Y$ is closed at order $\eps^2$ from \eqref{dj}. Hence, from \eqref{PPFLvaccuumdim4} we may be tempted to associate an internal energy $U^Y$ to the gravitational degrees of freedom such that its variation is $\Delta U^Y = \mathcal{Q}^Y = \frac{T_H}{8} \Delta A$. 

 \vspace{0.3 cm}

 To understand better the property of the gravitational flux and the application of the second law of thermodynamics on $\cN$, we generalize the previous analysis and write the balance law in the the $D$ dimensional spacetime. \footnote{In general $D$ dimensions, Sach's analysis of the free data on null hypersurfaces does not hold. We don't know what can be identified as gauge and what can identified as gravitational degrees of freedom in Bondi's frame.} In that case, \eqref{mastereqyork} becomes

\be 
    \Delta Q_\xi^Y = \frac{1}{16 \pi}\int_{\Delta \cN} \sigma_n^{\m \n} \pounds_\xi \g_{\m \n} \eps_\cN + 2 \frac{D - 3}{D - 2}\eps_\cN \pounds_\xi \theta_n + \int_{\Delta \cN} T_{\m \n} \xi^\m n^\n \eps_\cN \\ 
    \label{mastereqyork2}
\ee
and the charge is

\be
    Q_\xi^Y = \frac{1}{8 \pi} \int_S W(1 - \frac{\theta_{v \p_v}}{D - 2}) \eps_S - \frac{1}{2} P_A Y^A \eps_S
    \label{Yorkcharges}
\ee
If we restrict ourselves now to the null diffeomorphisms $\xi^\m = W n^\m$, \eqref{mastereqyork2} becomes

\be
    \Delta Q_{Wn}^Y = \frac{1}{8 \pi}\int_{\Delta \cN} W\eps_\cN \bigg( (D -3)[\frac{\theta_n}{\theta_{n0}} - 
    \big( \frac{\theta_n}{\theta_{n0}}\big)^2] + (1 - \frac{D - 3}{D -2}) \sigma_n^2 \bigg) + (1 - \frac{D - 3}{D - 2}) \int_{\Delta \cN} W T_{\m \n} n^\m n^\n \eps_\cN
    \label{positiveflux2}
\ee
with now $\theta_{n0} = D - 2$. The multiplicative factor $D - 3$ in front of the flux was expected because we know that we should not get any pure gravitational flux for $3$-dimensional gravity \footnote{Also there is no shear in $D = 3$}, as the Weyl tensor vanishes. \footnote{The number of gravitational degrees of freedom at each point in classical $D$ dimensional gravity is just $\frac{D(D+1)}{2} - 2D = \frac{D(D-3)}{2}$.} The analysis of the balance law on the $3$-dimensional light cone is held in Appendix.\ref{AppE}. We can also notice that as in the four dimensional case, the charge variation is positive as long as the null energy conditions are satisfied, we choose $W > 0$ and assume $0 \leq \theta_{n} \leq \theta_{n0}$. Now, from \eqref{positiveflux2}, we can get the first law for linearized perturbations around a stationary horizon. We can choose $W(x^A) = \kappa$ to be the surface gravity of the background stationary black hole in order to identify $\xi$ with the background Killing vector such that at infinity its timelike Killing component is normalized to $-1$. Thus, we recover the Hawking temperature $T_H$ in our formulas if we introduce the quantum of action $\hbar$. As we just did, let us specialize now to (local) vacuum, with $T_{\m \n} = 0$ on a portion of the horizon. Hence, at first order \eqref{positiveflux2} becomes

\be
    T_H \Delta S^Y = \Delta U^Y + O(\eps^4)
    \label{internalenergyvaria}
\ee
with $S^Y$ being the dynamical entropy

\be
    S^Y = \frac{k_B}{4 G \hbar}(A - \frac{v}{D-2}\frac{dA}{dv})
    \label{dynamicalYorkentropy}
\ee
and

\be
    U^Y = \frac{1}{2} k_B T_H \frac{D-3}{D-2} \frac{A}{2 G \hbar} = U^Y_{A \p_A}
    \label{internalenergy}
\ee
is analogous to an internal energy associated to the gravitational degrees of freedom, where $N = \frac{D-3}{D-2}\frac{A}{2 G \hbar}$ the number of independent gravitational degrees of freedom on a slice of constant $v$ if we assume equipartition (that is highly non trivial, mainly because we are studying charge variations between non equilibrium states) \footnote{Furthermore, equipartition only states that $U = \alpha N k_B T$ with $\alpha$ can take a large range of values. However, only if the Hamiltonian is quadratic in the configuration variable $q$, which basically means that it is an harmonic oscillator, we have $\alpha = \frac{1}{2}$}. This internal energy is similar to the one of a perfect gas with $N$ independent degrees of freedom. However, we should point out that we could also have identified the internal energy as \footnote{Indeed, remember that in vacuum at first order $\Delta A = \Delta (v \frac{dA}{dv})$ so there are several ways to remove the deltas.}

\be
    U^Y = \frac{1}{2} k_B T_H \frac{D-3}{D-2} \frac{A}{2 G \hbar} \frac{d \ln{A}}{d \ln{v}} = U^Y_{v \frac{\p}{\p v}}
    \label{internalenergy2}
\ee
that is \eqref{internalenergy} multiplied by the factor $\frac{d \ln{A}}{d \ln{v}}$. This factor can be interpreted as a kind of redshift. Indeed, the time generator $v \frac{\p}{\p v} = \frac{\p}{\p \ln{v}}$ associated to the labelling of the slices of $\cN$ by the normal $n = v \p_v$ is different from a more intrinsic "time" generator $A \frac{\p}{\p A} = \frac{\p}{\p \ln{A}}$ associated to the evolution of the geometry. As we know that the area of cross sections $A$ increases on the event horizon, the vector $A \p_A$ can potentially be thought as a time generator. Furthermore, near equilibrium, the dynamical physical configuration variable that we can identify from the York symplectic potential is the bare expansion $\bar{\theta}_n$. We can study the variations of this dynamical quantity in order to identify an intrinsic time scale. We check that

\be
    \frac{d \bar{\theta}_n}{d \ln{A}} = A \frac{d}{dA} (\frac{v}{A} \frac{dA}{dv}) = 1 - \frac{v}{A} \frac{dA}{dv} - v \frac{\frac{d^2 A}{dv^2}}{\frac{dA}{dv}} = 1 + O(\eps^2)
\ee
where we used the Raychaudhuri equation in vacuum in order to obtain the last equality. So $\ln{A}$ is the natural timescale associated to the dynamical event horizon near equilibrium (in the portions where there is no infalling matter). Thus, in vacuum, the two internal energies \eqref{internalenergy} and \eqref{internalenergy2} are associated to two different "time generators" $\frac{\p}{\p \ln{A}}$ and $\frac{\p}{\p \ln{v}}$ and related to each other through 

\be
    U^Y_{A \p_A} = \frac{d \ln{v}}{d \ln{A}} U^Y_{v \p_v}
\ee
but \eqref{internalenergy} is the internal energy constructed from the most physically relevant time generator on the near stationary event horizon. 

\subsubsection{Spherical symmetry and phase transition}

\begin{figure}[t]
\begin{center}
\includegraphics[scale=1.4]{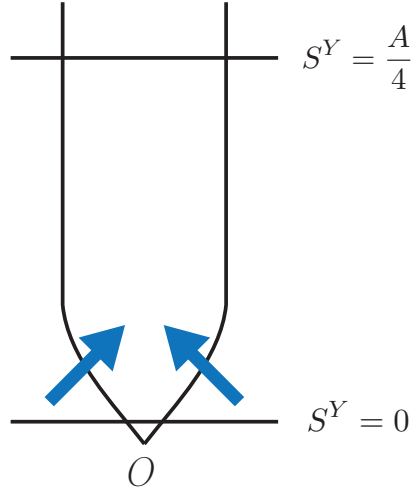}
\caption{Spherical symmetric collapse up to the formation of a black hole. In this case, the event horizon is a light cone "bent" by spacetime curvature once some matter entered it (blue arrows on the picture). This event horizon possesses only one caustic, at point $O$. The entropy $S^Y$ evolves from $0$ on the Minkowski light cone to $\frac{A}{4}$ once the black hole has reached its stationary state.}
\label{Figurelightcone}
\end{center}
\end{figure}

Let us consider an arbitrary point $O$ in spacetime and the bunch of light rays emanating from this point. Locally, spacetime is flat and so the bunch of outgoing light rays is a smooth submanifold which looks exactly like the outgoing light cone in Minkowski spacetime. As long as the topology of the bunch of light rays (if we remove $O$) is $S^{D-2} \times \mathbb{R}$ \footnote{This is not true in general for sufficiently large value of the affine parameter of the null geodesics}, we can use the analysis of the previous sections to define a local symmetry group and compute charges and fluxes. The Weyl supertranslation charge variation is given by \eqref{positiveflux2} 

\be
    \Delta Q_{Wn}^Y = \frac{1}{8 \pi}\int_{\Delta \cN} W\eps_\cN \bigg( (D -3)[\frac{\theta_n}{\theta_{n0}} - 
    \big( \frac{\theta_n}{\theta_{n0}}\big)^2] + \frac{1}{D -2} \sigma_n^2 \bigg) + \frac{1}{D - 2} \int_{\Delta \cN} W T_{\m \n} n^\m n^\n \eps_\cN
    \label{positiveflux3}
\ee
which is positive as long as the null energy conditions are satisfied, $W > 0$ and $\theta_{n0} = D-2 \geq \theta_n \geq 0$. Furthermore, the Raychaudhuri equation 
\be
    v \p_v \theta_n = \theta_n - \frac{1}{D - 2} \theta_n^2 - T_{\m \n} n^\m n^\m - \sigma_n^2
    \label{Raychaudhurieq}
\ee
tells us that if $\theta_n(v_0, x^A) = D - 2$, then $\theta_n (v > v_0, x^A) \leq D - 2$, exactly as in the four-dimensional case studied in section \ref{generalpropertyflux}. Hence, the analysis of this subsection extends well to the general $D$ dimensional case. However, if we pick some point $O$ in spacetime and look at the outgoing light ray emanating from this point, the expansion of this null congruence is $\theta_n(v, x^A) = \theta_n(0, x^A) = D - 2$ at least in a small neighborhood around $O$. Therefore, we have that $\theta_n(v > 0, x^A) \leq D - 2$.

\vspace{0.3 cm}

Let us assume now that spacetime is spherically symmetric about $O$. The null hypersurface spanned by the light rays emanating from $O$ is smooth and its topology is $S^{D-2} \times \mathbb{R}$. If the geodesic congruence is future complete, then the expansion $\theta_n$ is positive, and so the flux \eqref{positiveflux3} is positive and the charge $Q_n^Y$ increases. If a black hole forms, then $\theta_n$ ultimately reaches $0$ at infinity. If it is not the case, the expansion $\theta_n$ converges to $D - 2$ at when the affine parameter $v$ goes to infinity. Indeed, there must exist some parameter $v_0$ such that there is no matter or shear crossing the light cone for $v > v_0$. In that case, \eqref{Raychaudhurieq} reduces to 

\be
    v \p_v \theta_n = \theta_n - \frac{1}{D - 2} \theta_n^2
    \label{Raychaudhurieq2}
\ee
in the region $v > v_0$. As $0 < \theta_n \leq D - 2$, the rhs is positive. If it is the case,at infinity, we can prove that the charge $Q_n^Y$ becomes infinite if $D > 3$. This is because the gravitational flux does not fall off rapidly enough. On the contrary, on the event horizon, the flux \eqref{positiveflux3} vanishes near $v = 0$ and vanishes also for $v \rightarrow  + \infty$. Hence, the system evolves from the stationary state where the dynamical entropy \eqref{Yorkentropy} is $S^Y = 0$ to the stationary black hole state where $S^Y = \frac{A}{4}$ (see Figure \ref{Figurelightcone}). In that case, the parameter $\theta_n$ varies from $D - 2$ to $0$. For all the outgoing null hypersurfaces located at $u = u_0 < 0$, $\theta_n$ starts from $D - 2$ around $r = 0$, decreases when matter crosses $u = u_0$ (but never reaches $0$) before increasing again when matter stops falling and finally converges to $\theta_n = D - 2$. In other words

\be
    D - 2 = \underset{u\rightarrow0}{\lim} \underset{v\rightarrow+\infty}{\lim} \theta_n \neq \underset{v\rightarrow+\infty}{\lim} \underset{u\rightarrow0}{\lim} \theta_n = 0
\ee
These observations lead us to the conclusion that on the event horizon, the transition between the two equilibrium states where the flux and vanishes and the charges are constant is analogous to a phase transition with order parameter $\theta_n$. The phase of high symmetry is the stationary black hole. Indeed, while spherical symmetry is conserved during the whole process, this is not the case of time reversal symmetry, as in the "Minkowski's light cone phase", the area of the cross sections increases with increasing parameter $v$ \footnote{If we change $v$ into $u$, the expansion changes sign, so $\theta_n = -1$ as on the incoming light cone.} while on the stationary black hole the area of the cross sections does not. A naive way of understanding it is to remember that the physical properties of the flat light cone are not invariant if we shift $v$ to $v + v_0$ for arbitrary $v_0$, or $v$ to $e^\alpha v$ for arbitrary $\alpha$. However, the physical properties of the stationary black hole are invariant with respect to such shifts.

\vspace{0.3 cm}

In other words, the physical  properties of the spherically symmetric light cone are non invariant through (super) translations, while they are under rotations. Indeed, let consider the $D-1$ dimensional metric $q_{ab}$ \footnote{Here, we keep the Latin indices for tensors defined on codimension one manifolds, and the Greek indices for tensor defined on the $D$ dimensional spacetime. We note $q_{ab} = \pbi{g_{\m \n}}$} which is the pullback of $g_{\m \n}$ on the light cone. We know that the group

\be
    Diff(S^{D-2}) \ltimes \mathbb{R}_W^S
    \label{groupwt}
\ee
preserves the universal structure of the outgoing light cone, including the light cone boundary $O$. However, they do not all preserve the induced metric $q_{ab}$ on the light cone. Infinitesimal diffeomorphisms achieving this property satisfy 

\be
    \pounds_\xi q_{ab} = 0
    \label{lienull}
\ee
The generators of the rotation group satisfy \eqref{lienull}, as the light cone is spherically symmetric. However, all the Weyl supertranslations satisfy \eqref{lienull} if the expansion $\theta_n$ vanishes, and so the the transformations that leave physics invariant are the rotation group $SO(D-1)$ on the Minkowski light cone phase, where $\theta_n = D - 2$ while it is given by 

\be
    SO(D- 1) \ltimes \mathbb{R}_W^S
\ee
on the stationary black hole phase, in which $\theta_n = 0$. At a very late stage of the collapse, the black hole is similar to an eternal black hole. Therefore, as the affine supertranslations are local symmetries which preserve the background structure and satisfy \eqref{lienull}, they can be included in the symmetry group of the stationary event horizon. This is because the condition \eqref{lienull} is a local condition, so the fact that the light cone has a boundary in the past is irrelevant.  Thus, at very late time, the symmetry group of the stationary black hole is enlarged by all the supertranslations, and so it becomes
\be
    SO(D- 1) \ltimes \mathbb{R}_W^S \ltimes \mathbb{R}_T^S
\ee
All the supertranslations become part of the symmetry group if the order parameter $\theta_n$ we identified above vanishes. However, on the stationary black hole solution, the charges associated to the affine supertranslations vanish, while the charges associated to the Weyl supertranslations do not. In addition, the latter also vanish on the light cone solution \footnote{As we assume spherical symmetry, we expect that all the superrotation charges vanish during the whole process. In other words $P_A = 0$ in the coordinate system we chose adapted to the spherical symmetry. As there is no flux in the stationary phases, the charges vanish on any cross section.}. Hence, it suggests that the entropy going from $0$ on the flat light cone to $\frac{A}{4}$ on the stationary black hole might be a consequence of the appearance of new states labeled by the Weyl supertranslations charge aspect in a hypothetical Hilbert space. This observation relates to the seminal work of Hawking, Perry and Strominger on the role of supertranslations in order to solve the information loss paradox \cite{Hawking:2015qqa,Hawking:2016msc, Hawking:2016sgy}. On a non expanding horizon, these charges are given by 

\be
    Q_{Wn}^Y = \int_S W(x^A) \eps_S
    \label{generalsupercharge}
\ee
and so the charge aspect is the local area element $\eps_S$. However, for a stationary spherically symmetric black hole, we can decompose $W(x^A)$ into spherical harmonics. The charge \eqref{generalsupercharge} should vanish for all of them except for $l = m = 0$. Therefore there is no other charge than the total area. Nonetheless, even if a black hole forms from the spherical collapse of a gas cloud, the collapse is not perfectly spherically symmetric at all scales. It can be spherically symmetric on a macroscopic scale, such that the event horizon looks like a spherically symmetric light cone, but not on a microscopic level, because stress energy cannot be distributed homogeneously over all the angular directions \footnote{For instance, if the gas is a perfect gas made of non interacting atoms, the stress energy is focused on the location of the atoms and not between them. Of course, on a macroscopic scale, the stress energy is homogeneous.}. Therefore, the value of the charges \eqref{generalsupercharge} for higher-order spherical harmonics on the stationary phase can give information about the microscopic details of the gas that collapsed and formed the black hole. The precise relation between the charge variation and the stress energy flow across the event horizon is given by \eqref{positiveflux3}. 

\section{Outlook}
\label{outlook}

In this paper, we interpreted the master equation \eqref{master equation} contracted with a future null pointing diffeomorphism $\xi$ to a dynamical balance law for entropy, i.e a second law of thermodynamics. We discussed two possible choices of canonical flux and analyzed the properties of the associated thermodynamic potentials, i.e the dynamical entropies \eqref{Dirichletentropy} and \eqref{chargeyork}. In this framework, the entropy creation term is $T_{\m \n} \xi^\m n^\n \eps_\cN$, and is positive if the null energy conditions are satisfied.  It may open a discussion about the physical significance of the stress energy tensor. The following discussion is actually quite independent to the technical results obtained in this paper, but it was one of the main motivations to start this work, so it might be a good idea to talk a bit about these physical motivations at this stage. 

\vspace{0.3 cm}

The stress energy tensor $T_{\m \n}$ is often regarded as the covariant tensor associated to the energy density of matter, which is the source of the gravitational field and bends and distorts spacetime. However, if we interpret the master equation as a second law of thermodynamics, it might be relevant to think about $T_{\m \n}$ as a measure of entropy creation. Indeed, in non relativistic physics, energy is a conserved quantity associated with the time translation symmetry by Noether theorem, but the total energy of an isolated system can always be shifted without modification of the dynamics. In classical non-relativistic physics, it seems that in all physical principles that involve the energy of a physical system or subsystem, entropy maximisation is always the underlying fundamental principle. For instance, even if the Boltzmann factor depend explicitly on the energy of the subsystem, low energy states are favoured because they allow the reservoir to access a greater number of microstates. Similarly, in non relativistic quantum mechanics, the shift of the Hamiltonian only shifts the states of the system by an overall phase, with no incidence on the dynamics. However, the situation changes drastically in special relativity. Indeed, in this theory, space and time are merged in a subtle way, and so are space translation and time translation generators, i.e the momentum and the energy. As a consequence, energy becomes a measure of inertia (see \cite{Perez:2021iwh} for a very nice review about the equivalence between inertia and energy). In general relativity, the equivalence principle assures the equivalence between gravitational mass and inertia, and so between gravitational mass and energy. Hence energy is basically the source of gravitation, and indeed, we cannot "shift" the stress energy tensor by an "arbitrary constant" anymore, as we could do in non relativistic physics, because it is directly related to spacetime curvature.

\vspace{0.3 cm}

Of course, the exact meaning of energy in general relativity is intricate. As it has been reviewed in section.\ref{section2}, it is well known that the local stress energy tensor of a diffeomorphism invariant theory vanishes on-shell up to a boundary term, as the Euler-Lagrange equations are precisely the functional derivative of the Lagrangian with respect to the metric. Thereby, it is well known that the ADM or Bondi masses and angular momenta are charges defined through the introduction of an additional boundary structure at infinity. If the spacetime solution admits a Killing field, we can also defined conserved currents which can be interpreted as energy at infinity. However, as stressed out and discussed in section.\ref{section2}, the term $T_{\m \n} \xi^\m n^\n \eps_\cN$ must be interpreted as the black hole entropy variation at first order and not as the energy flux crossing the horizon. 

\vspace{0.3 cm}

Hence, if the interpretation of the balance law \ref{master equation} as an analog to a second law of thermodynamics is regarded as physically relevant, the "source" of gravity is nothing more that a dissipation term. Such an interpretation also implies that the positive energy conditions, in particular the null energy conditions, play a central role in order to understand gravity. Indeed, even if there is no assumption a priori for positive energy conditions in general relativity (or any other theory of gravity), it is well known that many theorems fail if they are not satisfied, in particular Hawking's classical area theorem \cite{Hawking:1971tu} and Penrose's singularity theorem \cite{Penrose:1964wq}. The null energy conditions are satisfied for non exotic classical matter, and  arguments have already been given to understand it as a consequence of gravity \cite{Parikh:2015ret}. However it is also well known that these positive energy conditions are violated for quantum matter \cite{Epstein:1965zza}, even if some physical quantities remain bounded. For instance, the average null energy condition on a null line remains true \cite{Wall:2009wi} and there exist inequalities analogous to the null energy conditions that are indeed satisfied at the quantum level \cite{Bousso:2015wca}. 

\section*{Acknowledgements}

I would like to thank Alejandro Perez, Manus Visser and Robert Wald for helpful discussions. I want especially to thank Simone Speziale for his teaching, the very valuable discussions we had about this paper, and for encouraging me to write it down. I would also like to thank an anonymous referee for useful comments.  

\appendix
\section{Electromagnetic balance law}
\label{AppA}

Here we derive a famous result, the energy conservation formula for electromagnetism, using the tools presented in section \ref{section2}. We use the same notations as in \ref{section2}. Here, the dynamical fields are the vector potential $A_\m$ and the background field is $g_{\m \n}$, such that $\d_\xi A_\m = \pounds_\xi A_\m$ and $\d g_{\m \n} = 0$. However, $\xi$ is a Killings field, i.e $\pounds_\xi g_{\m \n} = 0$. We start from \eqref{balancelawback} that we can now write as 
\begin{align}
    -d j_\xi &= \frac{\d L}{\d A_\m} \pounds_\xi A_\m \nn \\
    &= \frac{\d L}{\d A_\m} (i_\xi F)_\m + \frac{\d L}{\d A_\m} (d i_\xi A)_\m
    \label{dj3}
\end{align}
We can define $n$ through its relation to a volume form on the boundary of $\mathcal{M}$, such that we can write $\eps_\mathcal{M} = n \wedge \eps_{\p \mathcal{M}}$ and by integrating \eqref{dj3} over $\mathcal{M}$ we get
\begin{align}
    -\int_{\p \mathcal{M}} j_\xi &= \int_{\mathcal{M}} \frac{\d L}{\d A_\m} (i_\xi F)_\m + \frac{\d L}{\d A_\m} (d i_\xi A)_\m \nn \\
    \int_{\p \mathcal{M}} (\nabla_\n F^{\m \n} i_\xi A + T^\m_\n \xi^\n) n_\m \eps_{\p \mathcal{M}} &= \int_{\mathcal{M}} \frac{\d L}{\d A_\m} (i_\xi F)_\m + \nabla_\n F^{\m \n} \p_\m (i_\xi A) \eps_{\mathcal{M}} \nn \\
    \int_{\Sigma_2} T^\m_\n \xi^\n n_\m \eps_{\Sigma_2} - \int_{\Sigma_1} T^\m_\n \xi^\n n_\m \eps_{\Sigma_1} &= -\int_{\cN} T_{\m \n} \xi^\m n^\n \eps_{\cN} + \int_{\mathcal{M}} J^\m \xi^\n F_{\n \m} \eps_{\mathcal{M}}
\end{align}
where we used the Bianchi identity $\nabla_\m \nabla_\n F^{\m \n} = 0$ to get the third line. If $\xi = \frac{\p}{\p t}$ is a timelike Killing field, we get the famous balance law 
\be
    \Delta e = - \int_{\cN} \pi_i n^i \eps_{\cN} + \int_{\mathcal{M}} J^i E_i \eps_{\mathcal{M}}
\ee
where $\pi^i = (E \times B)^i$ in the Poynting vector, $E_i$ is the electric field, $J^i$ the three dimensional charged current and 
\be
    e = \f12(E^2 + B^2)
\ee
is the electromagnetic energy.

\section{Symmetry group on a null hypersurface at finite distance}
\label{AppB}

\subsection{Boundary structure preserving symmetry group} 

In this appendix, we aim to find the most general group of diffeomorphism such that

\begin{align}
\d_\xi n^\m = 0 \nn \\
\d_\xi n_\m = 0 \nn \\
\d_\xi k_n = 0
\label{conddef}
\end{align}
where $n^\m$ is a null normal of $\cN$ and $k_n$ being its unaffinity defined by $k_n = -l_\m n^\n \nabla_\n n^\m$. This symmetry group has already been found in \cite{chandrasekaran2018symmetries}, and is claimed to be the group which preserves the universal structure of the null hypersurface $\cN$, defined as the set of diffeomorphisms $\xi$ such that on $\cN$

\begin{align}
\pounds_\xi n^\m &= \beta n^\m \nn \\
\pounds_\xi n_\m &= \beta n_\m \nn \\
\pounds_\xi k_n &= \beta k_n + \pounds_n \beta
\label{groupdef}
\end{align}
It is shown in \cite{chandrasekaran2018symmetries} Appendix D that the diffeomorphisms satisfying \eqref{groupdef} also satisfy \eqref{conddef}, but the reverse is not explicitly worked up. Hence, as we started the discussion from the equations \eqref{conddef}, we derive the symmetry group which satisfies them. Following \cite{ashtekar2022charges} and Section.\ref{section4}, we work in a set of coordinates such that the null hypersurface $\cN$ is located at $u = 0$, and the affine parameter $v$ parameterizes the null geodesics on $\cN$. Hence, the vector $n = \frac{\p}{\p v}$ is tangent to the null geodesics and has vanishing unaffinity. As we study a (portion) of null hypersurface $\cN$ with topology $B \times \mathbb{R}$ we define the set set of coordinates $(u, v, x^A)$ in a neighborhood of $\cN$ such that a general metric in this neighborhood of $\cN$ can be written as \footnote{We see in \eqref{NUmetric} that $g_{vv} \stackrel{\cN}= O(u^2)$. This is needed because $v$ is an affine parameter. Indeed, a short calculation shows that $\frac{\p}{\p v}^\n \nabla_\n \frac{\p}{\p v}^\m = \Gamma^{\m}_{vv} = 0$ only if $\p_u g_{vv} \stackrel{\cN}= 0$.}

\be
    ds^2 = - u^2 F dv^2 + 2(-1 + uG) du dv + g_{uu} du^2 + 2 u P_A dx^A dv + 2 g_{uA} dx^A du + g_{AB} dx^A dx^B
    \label{NUmetric}
\ee
On $\cN$, \eqref{NUmetric} becomes

\be
    ds^2 \stackrel{u = 0}= - 2 du dv + g_{uu} du^2 + 2 g_{uA} du dx^A + g_{AB} dx^A dx^B
    \label{NUmetriconN}
\ee
Now, we see from \eqref{NUmetric} that the vector $n$ defined $n^\m = f (\frac{\p}{\p v})^\m$ with $f > 0$ is null, hypersurface orthogonal on $\cN$, and future-oriented. On $\cN$, its associated normal form is $n_\m = - f \p_\m u$. Furthermore, we can construct a vector $l$ such that $l^\m = \frac{1}{f} (\frac{\p}{\p u})^\m$, implying that

\be
    l^\m n_\m \stackrel{\cN}= -1
\ee
In this setup, we can take a closer look to the equations \eqref{conddef} and try to find out the infinitesimal diffeomorphisms $\xi$ satisfying them. By combining the first two equations of \eqref{conddef} we get on $\cN$

\be
    \pounds_\xi n^\m = g^{\m \n} \pounds_\xi n_\m
    \label{cond1contn}
\ee
Furthermore, as $\xi$ is tangent to $\cN$, the development of the $u$ component of $\xi$ around the null hypersurface $u = 0$ should be written as $\xi^u = - u W(v, x^A) + O(u^2)$. From this consideration, a short computation shows us that on $\cN$

\be
    \pounds_\xi n_\m = \omega_\xi n_\m
    \label{lienormform}
\ee
with

\be
    \omega_\xi = -l^\m \pounds_\xi n_\m = \xi^\n \p_\n f + \p_u \xi^u
    \label{omegaxi}
\ee
Hence, on $\cN$, $\pounds_\xi n^\m = \omega_\xi n^\m$ from \eqref{cond1contn} and so we get the following set of equation :

\begin{align}
    [\xi, n]^v &= \omega_\xi \nn \\
    [\xi, n]^A &= 0
    \label{eqsetlie}
\end{align}
The second equation of \eqref{eqsetlie} gives $\p_v \xi^A(v, x^B) = 0$, so

\be
    \xi^A(v, x^B) = \xi^A(x^B) = Y^A(x^B)
    \label{diffofsphere}
\ee
The infinitesimal diffeormorphisms \eqref{diffofsphere} are the linearizations of the diffeormorphisms of the $D-2$-sphere. Then, the first equation of \eqref{eqsetlie} can be re-written as :

\be
    \p_u \xi^u + \p_v \xi^v = 0
    \label{Weyleq}
\ee
and so

\be
    \xi^v(v, x^A) = \int_{v_0}^v W(v', x^A) dv'
    \label{xiv}
\ee
Now, we turn to the third equation of \eqref{conddef}. As on $\cN$ $\d_\xi n^\m = \d_\xi n_\m = 0$, we have $\d_\xi k_n = - n^\m l^\n \d_\xi \Gamma^\rho_{\mu \nu} n_\rho$, which gives the condition

\be
    \d_\xi k_n \stackrel{\cN}= l^\m \p_\m (n^\n n^\rho \pounds_\xi g_{\n \rho}) = 0
    \label{dksecondform}
\ee
As we still have $n^\m = f \p_v^\m$, \eqref{dksecondform} becomes

\begin{align}
    \d_\xi k_n &\stackrel{\cN}= \frac{1}{f}\p_u(f^2 \pounds_\xi g_{vv}) \nn \\
    &= -2f\p_vW(v, x^A) = 0
\end{align}
so $W(v, x^A) = W(x^A)$ and from \eqref{Weyleq} and \eqref{xiv} we deduce $\xi^v = T(x^A) + v W(x^A)$. Hence, the general linearized diffeomorphisms satisfying \eqref{conddef} are

\be
    \xi = (T(x^B) + v W(x^B)) \p_v - u W(x^B) \p_u + Y^A(x^B) \p_A
    \label{generaldiffeo}
\ee
in accordance with \cite{chandrasekaran2018symmetries}, where the components $T$ parameterize the affine supertranslations, the components $W$ the Weyl supertranslations, and the components $Y^A$ the superrotations. It also is the same group as the BMSW symmetry group at null infinity \cite{Freidel:2021fxf}, extending the famous BMS group. The bracket of two vectors $(T_1, W_1, Y_1^A)$ and $(T_2, W_2, Y_2^A)$ gives the following algebra \cite{chandrasekaran2018symmetries}

\begin{align}
    T_3 &= T_1 W_2 - T_2 W_1 + Y_1^A \p_A T_2 - Y_2^A \p_A T_1 \nn \\
    W_3 &= Y_1^A \p_A W_2 - Y_2^A \p_A W_1 \nn \\
    Y_3^A &= Y_1^B \p_B Y_2^A - Y_2^B \p_B Y_2^A
    \label{algebra}
\end{align}
The algebra is closed, and it is worth noticing that the subalgebras are also closed, in particular the subalgebra comprised of the vectors with $T = 0$. This symmetry group preserves the location of the corners of constant affine parameter $v$. Hence, if we consider surfaces $\cN$ with non trivial boundaries $\p \cN$, we should get rid of the affine supertranslations which move the boundaries and restrict ourselves to the symmetry group spanned by the Weyl supertranslations and the superrotations.

\subsection{Newman-Unti gauge}
However, in the main text we proceeded to a gauge fixing, and chose to work in the Newman-Unti gauge, as in \cite{ashtekar2022charges}. Hence we choose $g_{uu} = g_{uA} = 0$ and $g_{uv} = -1$ everywhere in \eqref{NUmetric}. If we make this choice, the vector $l^\m = \frac{1}{f} \frac{\p}{\p u}^\m$ is null, and is the auxiliary vector of $n^\m$ adapted to the foliation $v$. If we work with the metric \eqref{NUmetric0}, we also require that the diffeomorphisms $\xi$ preserves this gauge. Hence, in addition of \eqref{conddef}, we impose

\begin{align}
    \pounds_\xi g_{uu} &= 0 \implies \p_u \xi^v = 0 \nn \\
    \pounds_\xi g_{uA} &= 0 \implies \p_u \xi^A = g^{AB} \p_B \xi^v
    \label{NUgaugecons}
\end{align}
such that our diffeomorphisms of interest are

\be
    \xi = (T(x^B) + v W(x^B)) \p_v - u W(x^B) \p_u + [Y^A(x^B) +  u g^{AC} \p_C(T + vW)] \p_A + O(u^2)
    \label{generaldiffeo1}
\ee
as in  \cite{Chandrasekaran:2019ewn}. Hence, all the components of $\xi$ are fixed at first order. 

\section{Local temperature}
\label{AppC}

In this appendix, we define a notion of local temperature associated to some observers moving close to the null hypersurface along the lines drawn by the Weyl supertranslation field.  Indeed, choose a point $P$ close to $\cN$ \footnote{By close to $\cN$, we mean that the distance between one point of $\cN$ and $P$ is smaller the the curvature scale, such that there exists a small region of spacetime including a bunch of points of $\cN$ and $P$ in which the curvature effects are negligible.} with coordinates $(v_P, x_P^A)$ and consider the one dimensional curve defined by $x^A = x_P^A$, $v \geq v_P$, and $uv = \frac{1}{2W^2}$. The Weyl supertranslation vector field tangent to $H$ is given by \footnote{In the Rindler coordinate system in flat spacetime, we have $u = \frac{1}{\sqrt{2}W} e^{- W \tau}$ and $v = \frac{1}{\sqrt{2}W} e^{+ W \tau}$ where $\tau$ is the proper time of the accelerating observer with proper acceleration $W$. Hence at proper time $\tau$, she is at $(u(\tau), v(\tau)$, so the Killing boost tangent to the curve has components $\xi^u = \frac{du}{d\tau} = -Wu$ and $\xi^v = \frac{dv}{d\tau} = +Wv$. The difference between this case and the present analysis is that in flat spacetime $\xi$ is a global Killing field everywhere spanning an exact Killing horizon.}

\be
    \xi = W (v \p_v -  u \p_u)
    \label{monothermtrans}
\ee
has norm $\xi^\m \xi_\m = -1 + O(
\frac{1}{W^2})$ on $H$, and the norm of the acceleration vector field is given by

\be
    a^2 = g_{\mu \nu} \xi^\rho \nabla_\rho \xi^\m \xi^\sigma \nabla_\sigma \xi^\nu = -W^2 + O(W)
    \label{accelerationsquare}
\ee
In order to get these formulas, the acceleration $W$ must be large compared to the tidal forces and the gravitational twist, exact relations are given below. Locally, through the equivalence principle, the observer is uniformly accelerating with norm $W$ in flat spacetime with a local Killing horizon. If the acceleration $W$ is large compared the inverse proper time after which the curvature become non negligible, the observer accelerating along $H$ perceives a thermal spectrum \cite{barbado2012unruh, Parikh:2017aas}. Indeed, in quantum theory, the local Unruh temperature is

\be
    T_{ext} = \frac{W \hbar}{2 \pi c k_B}
    \label{exttemperature}
\ee
where the fundamental constants have been reintroduced on purpose. As $\frac{\hbar}{2 \pi c k_B}$ is very small, we need large acceleration in order to get sensitive Unruh temperature, consistently to the approximations we made in order to write $\xi^\m \xi_\m = -1$ and $a^\m a_\m = - W^2$ on $H$. This temperature should be measured by the thermometer  of an observer accelerating locally along $H$. However we could have chose another curve of constant local acceleration, and so the local temperature would be different. In any case, the quantity $W$ is not relevant in the local balance law  \eqref{Dirichletflux}, as it appears as a global multiplicative factor, and we can usually normalize it to one. Nevertheless, it gives a thermodynamic interpretation of the formula \eqref{Dirichletflux} from the point of view of an observer who is locally acceleration along $H$ and can carry a locally well defined notion of temperature, if the acceleration $W$ is large enough. We compute exact relations in the remaining of this section. 

\vspace{0.3 cm}

The metric near $\cN$ is given by \eqref{NUmetric0} and so the inverse metric is given by 

\begin{align}
    g^{-1} = u^2(g_{AB} P^A P^B + F) \p_u \p_u - 2 \p_u \p_v + 2 u P^A \p_u \p_A + g^{AB} \p_A \p_B
    \label{NUinversemetric}
\end{align}
where $P^A = g^{AB} P_B$. As we will be interested in constant external temperature, we will take $W(x^A) = W$ to simplify the calculations. The norm of $\xi$ on $H$ is given by

\be
    \xi^\m \xi_\mu = -1 - \frac{F}{4W^2}
    \label{normxi}
\ee
Our first assumption is to set $W^2 \gg F = -\frac{1}{2}\p_u^2 g_{vv} + o(u^2)$. Hence the acceleration must be much bigger than the local tidal force. However, in order to have measurable temperature, we need huge proper acceleration, of the order $\frac{c k_B}{\hbar}$ at least. Hence, this assumption seems to be relevant for our purpose, and for the following we will assume that $\xi_\n \xi^\n = -1$. The next step is to compute the acceleration vector given by

\be
    \xi^\n \nabla_\n \xi^\m = a^\m
    \label{acceleration}
\ee
The computation gives

\begin{align}
    a^v &= W^2 v (1 - \frac{F}{2W^2}) \nn \\
    a^u &= W^2 u(1 - \frac{1}{4W^2} \frac{\p F}{\p \ln{v}} + \frac{1}{2W^2} P^A \frac{\p P_A}{\p \ln{v}} + \frac{F}{W^2} - \frac{P^2}{2W^2} + \frac{1}{4W^3} - \frac{1}{4 W^4} P^A \p_A F)\nn \\
    a^A &= W^2 (\frac{P^A}{4 W^4} F + \frac{g^{AB}}{2W^2}\frac{\p P_B}{\p \ln{v}} + \frac{1}{8W^4} g^{AB} \p_B F - \frac{P^A}{W^2})
    \label{accelerationcoeff}
\end{align}
Then, the norm of the acceleration is given by

\begin{align}
    a^2 &= 2 g_{uv} a^v a^u + 2 g_{Av} a^v a^A + g_{vv} a^v a^v + g_{AB} a^A a^B \nn \\
    &= - W^2(1 - \frac{F}{2W^2})(1 - \frac{1}{4W^2} \frac{\p F}{\p \ln{v}} + \frac{1}{2W^2} P^A \frac{\p P_A}{\p \ln{v}} + \frac{F}{W^2} - \frac{P^2}{2W^2} + \frac{1}{4W^3} - \frac{1}{4 W^4} P^A \p_A F) \nn \\
    &+ W^2 (1 - \frac{F}{2 W^2})(\frac{P^2}{4 W^4} F + \frac{P^A}{2W^2}\frac{\p P_A}{\p \ln{v}} + \frac{1}{8W^4} P^A \p_A F - \frac{P^2}{W^2}) - u^2 v^2 W^2 (1 - \frac{F}{2W^2})F \nn \\
    &+ W^4 g_{AB}(\frac{P^A}{4 W^4} F + \frac{g^{AC}}{2W^2}\frac{\p P_C}{\p \ln{v}} + \frac{1}{8W^4} g^{AC} \p_C F - \frac{P^A}{W^2})(\frac{P^B}{4 W^4} F + \frac{g^{BC}}{2W^2}\frac{\p P_C}{\p \ln{v}} + \frac{1}{8W^4} g^{BC} \p_C F - \frac{P^B}{W^2}) \nn \\
    &= -W^2 + O(W)
\label{accelerationnorm}
\end{align}
The last equality of \eqref{accelerationnorm} makes sense only if the physical quantities $F = -\frac{1}{2} \p_u^2 g_{vv}$ and $P_A = \p_u g_{Av}$ verify $\frac{F}{W^2} \ll 1$ (the same condition as in \eqref{normxi} in order to have the norm of $\xi$ equal to $-1$) and $\frac{P_A}{W} \ll 1$. Furthermore, the derivatives of $F$ and $P_A$ with respect to the "time" $\ln{v}$ and the angular coordinates $A$ must also be very small compared to $W^2$ and $W$ respectively. Hence, we have to consider large enough accelerations, much larger than the local tidal forces and gravitational twist and their variations. However, as already noticed, we need significant accelerations in order to get a non infinitesimal Unruh temperature. If these conditions are satisfied, then \ref{accelerationnorm} gives us 

\be
    \vert a \vert = \sqrt{- a^2} = W
    \label{accapprox}
\ee
on $H$. Of course, \eqref{normxi} and \eqref{accapprox} are norms so they are invariant under a change of frame. Locally, the observer can always consider that spacetime is flat, and hence as he is submitted to constant acceleration $W$ for any point on $H$, and sees locally an Unruh temperature given by \eqref{exttemperature}

\section{Computation of anomalies}
\label{AppD}

Here we come back on the geometric quantities appearing in the flux \eqref{Dirichletflux} and \eqref{Yorkflux}. An analysis of the anomalies of the different physical quantities characterizing the intrinsic and extrinsic geometries of the null hypersurfaces already appears in \cite{chandrasekaran2021anomalies}, see also \cite{Odak2022}, and we will give a brief summary of the main results here. 
Let us consider a null hypersurface $\cN$ located at $u = 0$, with normal $n_\m = - f \p_\m u$ on $\cN$. If $\xi$ is tangent to $\cN$, we have

\be
    \pounds_\xi n_\m = \omega_\xi n_\m
    \label{formliederiv}
\ee
with 

\be
    \omega_\xi = \xi^\m \p_\m \ln{f} + \xi_1^u
    \label{proporfactor}
\ee
where $\xi^u = u \xi_1^u + O(u^2)$. We restrict ourselves further to the diffeomorphisms satisfying \eqref{groupdef} and preserving the universal structure of the null hypersurface. They correspond to an infinitesimal rescaling of the normal, and these are precisely the class III transformations \footnote{A class III transformation acts on a Newmann-Penrose null tetrad as $(n, l, m, \bar{m}) \rightarrow (An, A^{-1}l, me^{i \theta}, \bar{m} e^{-i\theta})$, where $n$ is the normal, $l$ an auxiliary vector while $m$ and its complex conjugate complete the basis.} from \cite{chandrasekhar1998mathematical}. Hence, a geometric quantity that is anomaly free must be class III invariant, because such quantities are invariant through a rescaling of the normal as in \eqref{eqclass}. It is straightforward to show from \eqref{formliederiv} and \eqref{volumeform}, \eqref{epsS}, \eqref{defexpansion}, \eqref{projector} and \ref{dictionary} that 

\begin{align}
    \Delta_\xi \eps_\cN &= \omega_\xi \eps_\cN \nn \\
    \Delta_\xi \eps_S &= 0 \nn \\
    \g^\alpha_\m \g^{\beta}_\n \Delta_\xi \gamma_{\alpha \beta} &= 0 \nn \\
    \Delta_\xi \theta &= -\omega_\xi \theta \nn \\
    \Delta_\xi k_n &= - \omega_\xi k_n - n^\m \p_\m \omega_\xi
    \label{anomalies}
\end{align}
Hence, the anomaly of all the relevant physical quantities on the null hypersurface depend linearly on $\omega_\xi$.  Now, we compute the anomalies associated to the diffeomorphisms \eqref{generaldiffeo0}. First, we need to compute the proportionality coefficient $\omega_\xi$ appearing in \eqref{lienormform}. Hence

\begin{equation}
\begin{aligned}
    \omega_\xi &= - l^\m \pounds_\xi n_\mu \\
    &= -W u \p_u \ln{f} + (T + v W) \p_v \ln{f} + Y^A \p_A \ln{f} - W
    \label{anomalyeq}
\end{aligned}
\end{equation}
We look for the diffeomorphisms $\xi$ such that $\omega_\xi = 0$. There are several interesting cases. If $T = W = 0$, then \eqref{anomalyeq} reduces to

\begin{equation}
    \omega_{Y^A \p_A} = Y^A \p_A \ln{f} = 0 
    \label{condrot}
\end{equation}
so if $f$ is independent of $x^B$ all the diffeomorphisms $\xi = Y^A \p_A$ are non anomalous. If $W = Y^A = 0$, then the anomalous diffeormorphisms are the one satisfying $\partial_v \ln{f} = 0$, so $f$ is independent on $v$. Third case, if $T = Y^A = 0$, then the equation becomes

\be
    v \p_v \ln{f} - u \p_u \ln{f}  = 1
    \label{condboost}
 \ee
The solutions of \ref{condboost} are given by

\begin{equation}
    f = c_1(x^A) v - \frac{c_2(x^A)}{u}
\end{equation}
but $c_2 = 0$ because $f$ must be defined on the null hypersurface $\cN$ located at $u =0$. Hence, as expected from \eqref{formliederiv}, the property of anomaly freedom does not rely only on the diffeomorphism $\xi$ but also on the chosen normal $n$. Therefore, when we choose the normal to be $n^\m = v (\frac{\p}{\p v})^\m$ as in section \ref{section4}, we have

\be
    \omega_{W (v \p v - u \p_u) + Y^A \p_A} = 0
\ee
but $\omega_{T \p_v} \neq 0$. We would have obtained a different result with another choice of normal, as $n^\m = \p_v^\m$ for instance, for which $\omega_{T \p_v + Y^A \p_A} = 0$ but $\omega_{W (v \p v - u \p_u)} \neq 0$. Hence, from \eqref{anomalies}, we understand that we can replace $\d_\xi \theta_n$ by $\pounds_\xi \theta_n$ for instance only for Weyl supertranslations and superrotations and disregarding the affine supertranslation, i.e we consider only the non anomalous diffeomorphisms

\be
    \xi = W (v \p_v - u \p_u) + Y^A(x^B) \p_A
\ee
in accordance to what we stated in the main text.

\section{Dynamical entropy of the 3D light cone}
\label{AppE}

This is an illustrative example of a system going through a succession of equilibrium states. Let consider the null hypersurface $\cN$ spanned by outgoing light rays starting from one point in flat spacetime in dimension $D = 3$. We still set $n^\m = v \p_v^\m$. It is well known that there exists no black hole solution in flat spacetime in dimension $D = 3$ because the Weyl tensor vanishes, even if such solutions exist for negative cosmological constant, as the BTZ black hole \cite{banados1992black, banados1993geometry}. However, we can still study the gravitational flux through $\cN$ and the gravitational charges cross sections. In dimension $D = 3$, there is no shear and so there is no gravitational flux, i.e $\Theta^D = \Theta^Y = 0$. Furthermore, the charges of both prescriptions are equal. We can compute the charges for a Weyl supertranslation $\xi = 2 \pi (v\p_v - u\p_u)$, and define in consequence the entropy

\be
    S = Q_\xi^D = Q_\xi^Y = \frac{1}{4}(A - v \frac{dA}{dv})
    \label{entropy3D}
\ee
The entropy variation on any portion of the null hypersurface $\cN$ is entirely given by the entropy creation term 

\be
    \frac{1}{2 \pi} \Delta S = \int_{\Delta \cN} T_{\m \n} \xi^\m n^\n \eps_\cN = \frac{1}{2 \pi} S_c
    \label{flux3D}
\ee
that is positive if the null energy conditions are imposed, as usual. Hence, on the outgoing light cone, the charge vanishes near $v = 0$, but increases as soon as some matter crosses it. During this process, spacetime is not flat. However, after some matter entered $\cN$, spacetime becomes flat again and the charges do not vary anymore. The entropy of the new stationary state is just given by

\be
    S = 2 \pi \int_{\Delta \cN} T_{\m \n} n^\m n^\n \eps_\cN \geq 0
    \label{newentropynew}
\ee
However, this non vanishing charge is not due the local geometry of the null hypersurface, as spacetime is flat in $D = 3$ in the absence of matter and cosmological constant. Hence, it accounts for the matter which crossed $\cN$ in the past. Hence, at any time $v$ at which the charge is stationary (no matter flux) the charge gives us the total matter flux that entered $\cN$ since $v = 0$, but does not give any precision on the history of the physical process.

\bibliographystyle{unsrt}
\bibliography{bibliographe.bib}

\begin{thebibliography}{10}

\bibitem{lee1990local}
Joohan Lee and Robert~M Wald.
\newblock Local symmetries and constraints.
\newblock {\em Journal of Mathematical Physics}, 31(3):725--743, 1990.

\bibitem{crnkovic1987covariant}
Cedomir Crnkovic and Edward Witten.
\newblock Covariant description of canonical formalism in geometrical theories.
\newblock {\em Three hundred years of gravitation}, pages 676--684, 1987.

\bibitem{ashtekar1991covariant}
Abhay Ashtekar, Luca Bombelli, and Oscar Reula.
\newblock The covariant phase space of asymptotically flat gravitational
  fields.
\newblock In {\em Mechanics, analysis and geometry: 200 years after Lagrange},
  pages 417--450. Elsevier, 1991.

\bibitem{iyer1994some}
Vivek Iyer and Robert~M Wald.
\newblock Some properties of the noether charge and a proposal for dynamical
  black hole entropy.
\newblock {\em Physical review D}, 50(2):846, 1994.

\bibitem{iyer1997lagrangian}
Vivek Iyer.
\newblock Lagrangian perfect fluids and black hole mechanics.
\newblock {\em Physical Review D}, 55(6):3411, 1997.

\bibitem{wald2000general}
Robert~M Wald and Andreas Zoupas.
\newblock General definition of ``conserved quantities'' in general relativity
  and other theories of gravity.
\newblock {\em Physical Review D}, 61(8):084027, 2000.

\bibitem{compere2019advanced}
Geoffrey Comp{\`e}re.
\newblock {\em Advanced lectures on general relativity}, volume 952.
\newblock Springer, 2019.

\bibitem{harlow2020covariant}
Daniel Harlow and Jie-qiang Wu.
\newblock Covariant phase space with boundaries.
\newblock {\em Journal of High Energy Physics}, 2020(10):1--52, 2020.

\bibitem{iyer1995comparison}
Vivek Iyer and Robert~M Wald.
\newblock Comparison of the noether charge and euclidean methods for computing
  the entropy of stationary black holes.
\newblock {\em Physical Review D}, 52(8):4430, 1995.

\bibitem{donnelly2016local}
William Donnelly and Laurent Freidel.
\newblock Local subsystems in gauge theory and gravity.
\newblock {\em Journal of High Energy Physics}, 2016(9):1--45, 2016.

\bibitem{speranza2018local}
Antony~J Speranza.
\newblock Local phase space and edge modes for diffeomorphism-invariant
  theories.
\newblock {\em Journal of High Energy Physics}, 2018(2):1--37, 2018.

\bibitem{freidel2020edge}
Laurent Freidel, Marc Geiller, and Daniele Pranzetti.
\newblock Edge modes of gravity. part i. corner potentials and charges.
\newblock {\em Journal of High Energy Physics}, 2020(11):1--52, 2020.

\bibitem{freidel2020edge2}
Laurent Freidel, Marc Geiller, and Daniele Pranzetti.
\newblock Edge modes of gravity. part ii. corner metric and lorentz charges.
\newblock {\em Journal of High Energy Physics}, 2020(11):1--64, 2020.

\bibitem{freidel2021edge}
Laurent Freidel, Marc Geiller, and Daniele Pranzetti.
\newblock Edge modes of gravity. part iii. corner simplicity constraints.
\newblock {\em Journal of High Energy Physics}, 2021(1):1--64, 2021.

\bibitem{freidel2021extended}
Laurent Freidel, Roberto Oliveri, Daniele Pranzetti, and Simone Speziale.
\newblock Extended corner symmetry, charge bracket and einstein's equations.
\newblock {\em Journal of High Energy Physics}, 2021(9):1--38, 2021.

\bibitem{chandrasekaran2021general}
Venkatesa Chandrasekaran, Eanna~E. Flanagan, Ibrahim Shehzad, and Antony~J.
  Speranza.
\newblock {A general framework for gravitational charges and holographic
  renormalization}.
\newblock {\em Int. J. Mod. Phys. A}, 37(17):2250105, 2022.

\bibitem{wald1993black}
Robert~M Wald.
\newblock Black hole entropy is the noether charge.
\newblock {\em Physical Review D}, 48(8):R3427, 1993.

\bibitem{jacobson1994black}
Ted Jacobson, Gungwon Kang, and Robert~C Myers.
\newblock On black hole entropy.
\newblock {\em Physical Review D}, 49(12):6587, 1994.

\bibitem{gao2001physical}
Sijie Gao and Robert~M Wald.
\newblock ``physical process version'' of the first law and the generalized
  second law for charged and rotating black holes.
\newblock {\em Physical Review D}, 64(8):084020, 2001.

\bibitem{gao2003first}
Sijie Gao.
\newblock First law of black hole mechanics in einstein-maxwell and
  einstein-yang-mills theories.
\newblock {\em Physical Review D}, 68(4):044016, 2003.

\bibitem{wald2018kerr}
Robert~M Wald.
\newblock Kerr--newman black holes cannot be over-charged or over-spun.
\newblock {\em International Journal of Modern Physics D}, 27(11):1843003,
  2018.

\bibitem{bardeen1973four}
James~M Bardeen, Brandon Carter, and Stephen~W Hawking.
\newblock The four laws of black hole mechanics.
\newblock {\em Communications in mathematical physics}, 31(2):161--170, 1973.

\bibitem{prabhu2017first}
Kartik Prabhu.
\newblock The first law of black hole mechanics for fields with internal gauge
  freedom.
\newblock {\em Classical and Quantum Gravity}, 34(3):035011, 2017.

\bibitem{hawking1972energy}
Stephen~W Hawking and JB~Hartle.
\newblock Energy and angular momentum flow into a black hole.
\newblock {\em Communications in mathematical physics}, 27(4):283--290, 1972.

\bibitem{wald1994quantum}
Robert~M Wald.
\newblock {\em Quantum field theory in curved spacetime and black hole
  thermodynamics}.
\newblock University of Chicago press, 1994.

\bibitem{Rignon-Bret:2023lyn}
Antoine Rignon-Bret.
\newblock {Note on the physical process first law of black hole mechanics}.
\newblock {\em Phys. Rev. D}, 108(2):024005, 2023.

\bibitem{mishra2018physical}
Akash Mishra, Sumanta Chakraborty, Avirup Ghosh, and Sudipta Sarkar.
\newblock On the physical process first law for dynamical black holes.
\newblock {\em Journal of High Energy Physics}, 2018(9):1--24, 2018.

\bibitem{sarkar2019black}
Sudipta Sarkar.
\newblock Black hole thermodynamics: general relativity and beyond.
\newblock {\em General Relativity and Gravitation}, 51:1--28, 2019.

\bibitem{ashtekar2002dynamical}
Abhay Ashtekar and Badri Krishnan.
\newblock Dynamical horizons: energy, angular momentum, fluxes, and balance
  laws.
\newblock {\em Physical review letters}, 89(26):261101, 2002.

\bibitem{ashtekar2004isolated}
Abhay Ashtekar and Badri Krishnan.
\newblock Isolated and dynamical horizons and their applications.
\newblock {\em Living Reviews in Relativity}, 7(1):1--91, 2004.

\bibitem{bousso2015new}
Raphael Bousso and Netta Engelhardt.
\newblock New area law in general relativity.
\newblock {\em Physical review letters}, 115(8):081301, 2015.

\bibitem{bousso2015proof}
Raphael Bousso and Netta Engelhardt.
\newblock Proof of a new area law in general relativity.
\newblock {\em Physical Review D}, 92(4):044031, 2015.

\bibitem{bhattacharjee2016entropy}
Srijit Bhattacharjee, Arpan Bhattacharyya, Sudipta Sarkar, and Aninda Sinha.
\newblock Entropy functionals and c-theorems from the second law.
\newblock {\em Physical Review D}, 93(10):104045, 2016.

\bibitem{Bhattacharjee:2015yaa}
Srijit Bhattacharjee, Sudipta Sarkar, and Aron~C. Wall.
\newblock {Holographic entropy increases in quadratic curvature gravity}.
\newblock {\em Phys. Rev. D}, 92(6):064006, 2015.

\bibitem{Chatterjee:2011wj}
Ayan Chatterjee and Sudipta Sarkar.
\newblock {Physical process first law and increase of horizon entropy for black
  holes in Einstein-Gauss-Bonnet gravity}.
\newblock {\em Phys. Rev. Lett.}, 108:091301, 2012.

\bibitem{Bhattacharya:2018xlq}
Krishnakanta Bhattacharya, Ashmita Das, and Bibhas~Ranjan Majhi.
\newblock {Noether and Abbott-Deser-Tekin conserved quantities in scalar-tensor
  theory of gravity both in Jordan and Einstein frames}.
\newblock {\em Phys. Rev. D}, 97(12):124013, 2018.

\bibitem{Dey:2021rke}
Sumit Dey, Krishnakanta Bhattacharya, and Bibhas~Ranjan Majhi.
\newblock {Thermodynamic structure of a generic null surface and the zeroth law
  in scalar-tensor theory}.
\newblock {\em Phys. Rev. D}, 104(12):124038, 2021.

\bibitem{Bhattacharya:2022mnb}
Krishnakanta Bhattacharya and Bibhas~Ranjan Majhi.
\newblock {Scalar tensor gravity from thermodynamic and fluid-gravity
  perspective}.
\newblock {\em Gen. Rel. Grav.}, 54(9):112, 2022.

\bibitem{chakraborty2015thermodynamical}
Sumanta Chakraborty and T~Padmanabhan.
\newblock Thermodynamical interpretation of the geometrical variables
  associated with null surfaces.
\newblock {\em Physical Review D}, 92(10):104011, 2015.

\bibitem{chakraborty2015gravitational}
Sumanta Chakraborty, Krishnamohan Parattu, and T~Padmanabhan.
\newblock Gravitational field equations near an arbitrary null surface
  expressed as a thermodynamic identity.
\newblock {\em Journal of High Energy Physics}, 2015(10):1--27, 2015.

\bibitem{adami2022null}
H~Adami, MM~Sheikh-Jabbari, V~Taghiloo, and H~Yavartanoo.
\newblock Null surface thermodynamics.
\newblock {\em Physical Review D}, 105(6):066004, 2022.

\bibitem{dey2020covariant}
Sumit Dey and Bibhas~Ranjan Majhi.
\newblock Covariant approach to the thermodynamic structure of a generic null
  surface.
\newblock {\em Physical Review D}, 102(12):124044, 2020.

\bibitem{Duval:2014uva}
Christian Duval, Gary~W. Gibbons, and Peter~A. Horvathy.
\newblock {Conformal Carroll groups and BMS symmetry}.
\newblock {\em Class. Quant. Grav.}, 31:092001, 2014.

\bibitem{Duval:2014lpa}
Christian Duval, Gary~W. Gibbons, and Peter~A. Horvathy.
\newblock {Conformal Carroll groups}.
\newblock {\em J. Phys. A}, 47(33):335204, 2014.

\bibitem{chandrasekaran2018symmetries}
Venkatesa Chandrasekaran, {\'E}anna~{\'E} Flanagan, and Kartik Prabhu.
\newblock Symmetries and charges of general relativity at null boundaries.
\newblock {\em Journal of High Energy Physics}, 2018(11):1--68, 2018.

\bibitem{Ciambelli:2018wre}
Luca Ciambelli, Charles Marteau, Anastasios~C. Petkou, P.~Marios Petropoulos,
  and Konstantinos Siampos.
\newblock {Flat holography and Carrollian fluids}.
\newblock {\em JHEP}, 07:165, 2018.

\bibitem{Ciambelli:2019lap}
Luca Ciambelli, Robert~G. Leigh, Charles Marteau, and P.~Marios Petropoulos.
\newblock {Carroll Structures, Null Geometry and Conformal Isometries}.
\newblock {\em Phys. Rev. D}, 100(4):046010, 2019.

\bibitem{Donnay:2019jiz}
Laura Donnay and Charles Marteau.
\newblock {Carrollian Physics at the Black Hole Horizon}.
\newblock {\em Class. Quant. Grav.}, 36(16):165002, 2019.

\bibitem{hopfmuller2017gravity}
Florian Hopfm{\"u}ller and Laurent Freidel.
\newblock Gravity degrees of freedom on a null surface.
\newblock {\em Physical Review D}, 95(10):104006, 2017.

\bibitem{hopfmuller2018null}
Florian Hopfm{\"u}ller and Laurent Freidel.
\newblock Null conservation laws for gravity.
\newblock {\em Physical Review D}, 97(12):124029, 2018.

\bibitem{Grumiller:2019fmp}
Daniel Grumiller, Alfredo P\'erez, M.~M. Sheikh-Jabbari, Ricardo Troncoso, and
  C\'eline Zwikel.
\newblock {Spacetime structure near generic horizons and soft hair}.
\newblock {\em Phys. Rev. Lett.}, 124(4):041601, 2020.

\bibitem{oliveri2020boundary}
Roberto Oliveri and Simone Speziale.
\newblock Boundary effects in general relativity with tetrad variables.
\newblock {\em General Relativity and Gravitation}, 52(8):1--52, 2020.

\bibitem{ashtekar2022non}
Abhay Ashtekar, Neev Khera, Maciej Kolanowski, and Jerzy Lewandowski.
\newblock Non-expanding horizons: multipoles and the symmetry group.
\newblock {\em Journal of High Energy Physics}, 2022(1):1--33, 2022.

\bibitem{ashtekar2022charges}
Abhay Ashtekar, Neev Khera, Maciej Kolanowski, and Jerzy Lewandowski.
\newblock Charges and fluxes on (perturbed) non-expanding horizons.
\newblock {\em Journal of High Energy Physics}, 2022(2):1--38, 2022.

\bibitem{Odak2022}
Gloria Odak, Antoine Rignon-Bret, and Simone Speziale.
\newblock {\em in preparation}, 2023.

\bibitem{jacobson1995thermodynamics}
Ted Jacobson.
\newblock Thermodynamics of spacetime: the einstein equation of state.
\newblock {\em Physical Review Letters}, 75(7):1260, 1995.

\bibitem{Shi:2020csw}
Kai Shi, Xuan Wang, Yihong Xiu, and Hongbao Zhang.
\newblock {Covariant phase space with null boundaries}.
\newblock {\em Commun. Theor. Phys.}, 73(12):125401, 2021.

\bibitem{Ciambelli:2021nmv}
Luca Ciambelli, Robert~G. Leigh, and Pin-Chun Pai.
\newblock {Embeddings and Integrable Charges for Extended Corner Symmetry}.
\newblock {\em Phys. Rev. Lett.}, 128, 2022.

\bibitem{Ciambelli:2022cfr}
Luca Ciambelli and Robert~G. Leigh.
\newblock {Universal corner symmetry and the orbit method for gravity}.
\newblock {\em Nucl. Phys. B}, 986:116053, 2023.

\bibitem{Freidel:2023bnj}
Laurent Freidel, Marc Geiller, and Wolfgang Wieland.
\newblock {Corner symmetry and quantum geometry}.
\newblock 2 2023.

\bibitem{Ciambelli:2022vot}
Luca Ciambelli.
\newblock {From Asymptotic Symmetries to the Corner Proposal}.
\newblock {\em PoS}, Modave2022:002, 2023.

\bibitem{Hawking:1971tu}
Stephen~W. Hawking.
\newblock {Gravitational radiation from colliding black holes}.
\newblock {\em Phys. Rev. Lett.}, 26:1344--1346, 1971.

\bibitem{Wall:2015raa}
Aron~C. Wall.
\newblock {A Second Law for Higher Curvature Gravity}.
\newblock {\em Int. J. Mod. Phys. D}, 24(12):1544014, 2015.

\bibitem{Wall:2011hj}
Aron~C. Wall.
\newblock {A proof of the generalized second law for rapidly changing fields
  and arbitrary horizon slices}.
\newblock {\em Phys. Rev. D}, 85:104049, 2012.
\newblock [Erratum: Phys.Rev.D 87, 069904 (2013)].

\bibitem{Parikh:2015ret}
Maulik Parikh and Andrew Svesko.
\newblock {Thermodynamic Origin of the Null Energy Condition}.
\newblock {\em Phys. Rev. D}, 95(10):104002, 2017.

\bibitem{Freidel:2021fxf}
Laurent Freidel, Roberto Oliveri, Daniele Pranzetti, and Simone Speziale.
\newblock {The Weyl BMS group and Einstein\textquoteright{}s equations}.
\newblock {\em JHEP}, 07:170, 2021.

\bibitem{Wald2023}
Robert Wald and Victor Zhang.
\newblock {\em in preparation}, 2023.

\bibitem{Visser2023}
Manus Visser and Zihan Yan.
\newblock {Dynamical Black Hole Entropy is Improved Noether Charge}.
\newblock {\em in preparation}, 2023.

\bibitem{Odak:2022ndm}
Gloria Odak, Antoine Rignon-Bret, and Simone Speziale.
\newblock {Wald-Zoupas prescription with (soft) anomalies}.
\newblock 12 2022.

\bibitem{wald1990identically}
Robert~M Wald.
\newblock On identically closed forms locally constructed from a field.
\newblock {\em Journal of mathematical physics}, 31(10):2378--2384, 1990.

\bibitem{noether1971invariant}
Emmy Noether.
\newblock Invariant variation problems.
\newblock {\em Transport theory and statistical physics}, 1(3):186--207, 1971.

\bibitem{parattu2016boundary}
Krishnamohan Parattu, Sumanta Chakraborty, Bibhas~Ranjan Majhi, and
  T~Padmanabhan.
\newblock A boundary term for the gravitational action with null boundaries.
\newblock {\em General Relativity and Gravitation}, 48(7):1--28, 2016.

\bibitem{lehner2016gravitational}
Luis Lehner, Robert~C Myers, Eric Poisson, and Rafael~D Sorkin.
\newblock Gravitational action with null boundaries.
\newblock {\em Physical Review D}, 94(8):084046, 2016.

\bibitem{DePaoli:2018erh}
Elena De~Paoli and Simone Speziale.
\newblock {A gauge-invariant symplectic potential for tetrad general
  relativity}.
\newblock {\em JHEP}, 07:040, 2018.

\bibitem{Oliveri:2020xls}
Roberto Oliveri and Simone Speziale.
\newblock {A note on dual gravitational charges}.
\newblock {\em JHEP}, 12:079, 2020.

\bibitem{sachs1962asymptotic}
Rainer~K Sachs.
\newblock Asymptotic symmetries in gravitational theory.
\newblock {\em Physical Review}, 128(6):2851, 1962.

\bibitem{sachs1962characteristic}
Rainer~K Sachs.
\newblock On the characteristic initial value problem in gravitational theory.
\newblock {\em Journal of Mathematical Physics}, 3(5):908--914, 1962.

\bibitem{Hawking:2015qqa}
Stephen~W. Hawking.
\newblock {The Information Paradox for Black Holes}.
\newblock 9 2015.

\bibitem{Hawking:2016msc}
Stephen~W. Hawking, Malcolm~J. Perry, and Andrew Strominger.
\newblock {Soft Hair on Black Holes}.
\newblock {\em Phys. Rev. Lett.}, 116(23):231301, 2016.

\bibitem{Hawking:2016sgy}
Stephen~W. Hawking, Malcolm~J. Perry, and Andrew Strominger.
\newblock {Superrotation Charge and Supertranslation Hair on Black Holes}.
\newblock {\em JHEP}, 05:161, 2017.

\bibitem{chandrasekhar1998mathematical}
Subrahmanyan Chandrasekhar.
\newblock {\em The mathematical theory of black holes}, volume~69.
\newblock Oxford university press, 1998.

\bibitem{Chandrasekaran:2019ewn}
Venkatesa Chandrasekaran and Kartik Prabhu.
\newblock {Symmetries, charges and conservation laws at causal diamonds in
  general relativity}.
\newblock {\em JHEP}, 10:229, 2019.

\bibitem{de2018light}
Tommaso De~Lorenzo and Alejandro Perez.
\newblock Light cone thermodynamics.
\newblock {\em Physical Review D}, 97(4):044052, 2018.

\bibitem{de2019light}
Tommaso De~Lorenzo and Alejandro Perez.
\newblock Light cone black holes.
\newblock {\em Physical Review D}, 99(6):065009, 2019.

\bibitem{Barnich:2009se}
Glenn Barnich and Cedric Troessaert.
\newblock {Symmetries of asymptotically flat 4 dimensional spacetimes at null
  infinity revisited}.
\newblock {\em Phys. Rev. Lett.}, 105:111103, 2010.

\bibitem{Barnich:2010eb}
Glenn Barnich and Cedric Troessaert.
\newblock {Aspects of the BMS/CFT correspondence}.
\newblock {\em JHEP}, 05:062, 2010.

\bibitem{Barnich:2011mi}
Glenn Barnich and Cedric Troessaert.
\newblock {BMS charge algebra}.
\newblock {\em JHEP}, 12:105, 2011.

\bibitem{Perez:2021iwh}
Alejandro Perez and Salvatore Ribisi.
\newblock {Energy-mass equivalence from Maxwell equations}.
\newblock {\em Am. J. Phys.}, 90(4):305, 2022.

\bibitem{Penrose:1964wq}
Roger Penrose.
\newblock {Gravitational collapse and space-time singularities}.
\newblock {\em Phys. Rev. Lett.}, 14:57--59, 1965.

\bibitem{Epstein:1965zza}
H.~Epstein, V.~Glaser, and A.~Jaffe.
\newblock {Nonpositivity of energy density in Quantized field theories}.
\newblock {\em Nuovo Cim.}, 36:1016, 1965.

\bibitem{Wall:2009wi}
Aron~C. Wall.
\newblock {Proving the Achronal Averaged Null Energy Condition from the
  Generalized Second Law}.
\newblock {\em Phys. Rev. D}, 81:024038, 2010.

\bibitem{Bousso:2015wca}
Raphael Bousso, Zachary Fisher, Jason Koeller, Stefan Leichenauer, and Aron~C.
  Wall.
\newblock {Proof of the Quantum Null Energy Condition}.
\newblock {\em Phys. Rev. D}, 93(2):024017, 2016.

\bibitem{barbado2012unruh}
Luis~C Barbado and Matt Visser.
\newblock Unruh-dewitt detector event rate for trajectories with time-dependent
  acceleration.
\newblock {\em Physical Review D}, 86(8):084011, 2012.

\bibitem{Parikh:2017aas}
Maulik Parikh and Andrew Svesko.
\newblock {Einstein\textquoteright{}s equations from the stretched future light
  cone}.
\newblock {\em Phys. Rev. D}, 98(2):026018, 2018.

\bibitem{chandrasekaran2021anomalies}
Venkatesa Chandrasekaran and Antony~J Speranza.
\newblock Anomalies in gravitational charge algebras of null boundaries and
  black hole entropy.
\newblock {\em Journal of High Energy Physics}, 2021(1):1--56, 2021.

\bibitem{banados1992black}
Maximo Banados, Claudio Teitelboim, and Jorge Zanelli.
\newblock Black hole in three-dimensional spacetime.
\newblock {\em Physical Review Letters}, 69(13):1849, 1992.

\bibitem{banados1993geometry}
Maximo Banados, Marc Henneaux, Claudio Teitelboim, and Jorge Zanelli.
\newblock Geometry of the 2+ 1 black hole.
\newblock {\em Physical Review D}, 48(4):1506, 1993.

\end{thebibliography}
\end{document}